\documentclass[12pt,a4paper]{article} 
\pdfoutput=1
\usepackage{jheppub}
\usepackage{enumerate}
\usepackage{epsfig} 
\usepackage{float} 
\usepackage[caption = false]{subfig}
\usepackage{wasysym} 
\usepackage{mathrsfs} 
\usepackage{amsfonts} 
\usepackage{amsbsy} 
\usepackage{amscd} 
\usepackage{pstricks} 
\usepackage{multirow} 
\usepackage{tikz}
\usepackage{color}
\usepackage{slashed}
\usepackage{array}
\usepackage{tablefootnote}

\usetikzlibrary{arrows,positioning,shapes.geometric} 
\usepackage{color}
\definecolor{myred}{rgb}{0.6,0,0} %usage:  {\textcolor{myred}{Hello World}}
\definecolor{myblue}{rgb}{0,0.2,0.4}
\definecolor{mygreen}{rgb}{0,0.9,0.1}
\definecolor{hc}{rgb}{.9,0.1,0.7}
\definecolor{hcout}{rgb}{.9,0.7,0.9}
\definecolor{Orange}{rgb}{1.,0.65,0.}
\numberwithin{equation}{section}
\numberwithin{figure}{section}
\numberwithin{table}{section}

\newcommand{\met}{\ensuremath{\not\!\!E_T}\xspace}
\newcommand{\be}{\begin{equation}}
\newcommand{\ee}{\end{equation}}
\newcommand{\bea}{\begin{eqnarray}}
\newcommand{\eea}{\end{eqnarray}}

\newcommand{\newc}{\newcommand}
\newc{\bi}{\begin{itemize}}
\newc{\ei}{\end{itemize}}

\newc{\ra}{\rightarrow}
\newc{\sq}   {\mbox{$\wt{q}$}}
\newc{\msq}  {\mbox{$m_{\sq}$}}
\newc{\gl}   {\mbox{$\wt{g}$}}
\newc{\mgl}  {\mbox{$m_{\gl}$}}
\def \met  {\mbox{${E\!\!\!\!/_T}$}}
\newc{\wt}{\widetilde}

\def \lspone{\wt\chi_1^0}
\def \mlspone{m_{\lspone}}

\newc{\ifb}{\mbox{${\rm fb}^{-1}$}}

\newc{\del}{\delta}

\def \hstop{\wt {t}_2}

\def \hsbot{\wt {b}_2}

\def \grav{\wt {G}}
\def \mgrav{m_{\grav}}

\title{Search for a compressed supersymmetric spectrum with a light Gravitino} 
\author[a]{Juhi Dutta,}  
\author[b]{Partha Konar,}
\author[a]{Subhadeep Mondal,}
\author[a]{Biswarup Mukhopadhyaya,}
\author[a]{Santosh Kumar Rai}
\affiliation[a]{ Regional Centre for Accelerator-based Particle Physics, \\
Harish-Chandra Research Institute, HBNI, \\ Chhatnag Road, Jhusi, Allahabad 211019, India}
\affiliation[b]{Physical Research Laboratory,  Ahmedabad 380009, India}
\emailAdd{juhidutta@hri.res.in}
\emailAdd{konar@prl.res.in}
\emailAdd{subhadeepmondal@hri.res.in}
\emailAdd{biswarup@hri.res.in}
\emailAdd{skrai@hri.res.in}

\abstract{
Presence of the light 
gravitino as  dark matter candidate in a supersymmetric (SUSY) model opens up interesting collider signatures consisting of one or more 
hard photons together with multiple jets and missing transverse energy from the cascade decay. 
We investigate such signals at the 13 TeV LHC in presence 
of compressed SUSY spectra, consistent with the Higgs mass as well as collider and dark matter constraints. We analyse and compare the 
discovery potential in different benchmark scenarios consisting of both compressed and uncompressed SUSY spectra, considering 
different levels of compression and intermediate decay modes.  Our conclusion is that compressed spectra upto 2.5 TeV are likely to be probed even before the 
high luminosity run of LHC. Kinematic variables are also suggested, which offer distinction between compressed and uncompressed spectra
yielding similar event rates for photons + multi-jets + $\met$. 
}

\preprint{HRI-RECAPP-2017-005}
\begin{document}

\maketitle
%%%%%%%%%%%%%%%%%%%%%%
\section{Introduction}
\label{sec:into}
%%%%%%%%%%%%%%%%%%%%%%
The Large Hadron Collider (LHC) has already accumulated a substantial volume of data with $\sqrt{s}=13$ TeV. 
Although the discovery of a scalar resembling the Higgs boson \cite{Aad:2015zhl,ATLAS-CONF-2016-081,Aad:2012tfa,Chatrchyan:2012xdj,CMS-PAS-HIG-16-020,
ATLAS-CONF-2016-067} in the Standard Model (SM) has laid the foundation 
of a success story, the absence of any new physics signal is a source of exasperation to those in search of physics 
beyond the SM (BSM). This applies to the search for phenomenologically viable supersymmetric (SUSY) 
scenarios as well. The non-observation of any supersymmetric particle so far at the LHC has strengthened the limits on many 
such low scale SUSY models. While the large production 
cross-section of the coloured SUSY particles (sparticles) are already pushing the existing mass limits to the 2 TeV mark 
with the initial data at the 13 TeV run, the weakly interacting sparticles are still not that severely constrained\cite{MoriondATLAS,MoriondCMS}. With the 
LHC already operating close to its near maximum centre-of-mass energy,  consistent improvements in luminosity is 
expected to help accumulate enough data which will help probe the coloured sector mass to almost 3 TeV with some 
improvements for the weakly interacting sector too. 

This lack of evidence for any low scale SUSY events prompted the idea of a compressed sparticle 
spectrum \cite{Martin:2008aw,LeCompte:2011cn,LeCompte:2011fh,Alvarez:2012wf,Bhattacherjee:2012mz,Belanger:2012mk,Dreiner:2012sh,
Bhattacherjee:2013wna,Mukhopadhyay:2014dsa,Dutta:2015exw,Chakraborty:2016qim,Chowdhury:2016qnz,Nath:2016kfp}, where 
the lightest SUSY particle (LSP) and  the heavier sparticle states may be nearly degenerate. In such realizations 
of the mass spectra, the resulting final state jets and leptons from the decay cascades of the parent particles are expected to 
be very soft, including the overall missing transverse energy which is a manifestation of the available visible transverse momenta. 
As events with such soft final states would be susceptible to low acceptance efficiencies in the detectors and therefore lead to 
much smaller event rates in the conventional SUSY search channels.
In the absence of hard leptons or jets arising from the cascade, one has to rely on  tagging 
the jets or photons originating from the initial state radiation (ISR) or final state radiation (FSR) to detect such events where
the available missing transverse momenta is characterized by the stability of the LSP in the cascades. Usually, in most SUSY models, the lightest neutralino ($\lspone$) is assumed to be the LSP. 
Thus, such signals allow a much lighter SUSY spectrum compared to the conventional channels with hard leptons, jets and large 
missing transverse momentum \cite{Chatrchyan:2011nd,Aad:2011xw,Aad:2012fqa,Aad:2014wea,Chatrchyan:2014lfa, 
Khachatryan:2014rra,PhysRevD.94.032005,Aad:2015zva, Aaboud:2016zdn,ATLAS:2016kts,CMS:2016nnn}.

However, in the presence of a light gravitino ($\grav$) in the spectrum, such as in gauge mediated SUSY breaking (GMSB) 
models \cite{Dimopoulos:1996vz,Zwirner:1997gt,Giudice:1998bp,Feng:2010ij,Ruderman:2011vv,Kim:2017pvm,Dreiner:2011fp,Allanach:2016pam},
the $\lspone$ is quite often the next-to-lightest SUSY particle (NLSP), which decays into a $\grav$ and a gauge/Higgs boson. 
Search strategy for such scenarios, therefore, is expected to be significantly different. In this case, one would always expect to 
find one or more hard leptons/jets/photons in the final state originating from the $\lspone$ decay, 
irrespective of whether the SUSY mass spectrum is compressed or not.
Hence detecting events characterizing such a signal is expected to be much easier, with the preferred channel 
being the photon mode. Given the fact that the hard photon(s) can easily be tagged for these events in a relatively compressed
spectrum of the SUSY particles with the NLSP, one need not rely on the radiated jets for signal identification,
thereby improving the cut efficiency significantly. If one considers a fixed gravitino mass, the photon(s) originating from the 
$\lspone$ decay will be harder as $\mlspone$ becomes heavier. Hence these hard photon associated signals can be 
very effective to probe a heavy SUSY spectrum with a light gravitino as there would rarely be any SM events with such 
hard photons in the final state. 

While the light gravitino scenario yields large transverse missing energy ($\met$) as well as hard photon(s) and jet(s), 
the question remains as to whether its presence obliterate the information on whether the MSSM part of the spectrum is compressed 
or not. In this work, we have demonstrated how such information can be extracted. 
Our study in this direction contains the following new observations:
\begin{itemize}
\item A set of kinematic observables are identified involving hardness of the photon(s), the transverse momenta ($p_T$) of the 
leading jets and also the $\met$, which clearly brings out the distinction between a compressed and an uncompressed spectrum with similar signal rates.
We have studied different benchmarks with varied degree of compression in the spectrum in this context. 

\item The characteristic rates of the $n$-$\gamma$ (where $n\ge 1$) final state in a compressed spectrum scenario have been 
obtained and the underlying physics has been discussed.
\item The circumstances under which, for example, a gluino in a compressed MSSM spectrum prefers to decay into a gluon and a gravitino 
rather than into jets and a neutralino have been identified. In this context, we have also found some remarkable effects of a eV-scale 
gravitino though such a particle can not explain the cold dark matter (DM) content of the universe.
\end{itemize}

The experimental collaborations have considered light gravitino scenarios and derived bounds on the coloured sparticles
\cite{Aad:2012zza,Aad:2015hea,CMS:2016igk,Khachatryan:2016hns,CMS:2016nij,ATLASCollaboration:2016wlb,Khachatryan:2016ojf,
ATLAS:2016fks}. 
The ATLAS collaboration recently published their analysis on a SUSY scenario with a light $\grav$ with the 13 TeV data accumulated at 
an integrated luminosity of $13.3~{\rm fb^{-1}}$ \cite{ATLAS:2016fks}. In this analysis, $\lspone$ is considered to be a bino-higgsino 
mixed state decaying into  $\gamma\grav$ and(or) $Z\grav$ resulting in the final state  
``$n_1 \, \gamma + n_2$ jets + $\met$''  where $n_1\ge 1$ and $n_2 > 2$. The 13 TeV data puts a stringent 
constraint on the sparticle masses excluding $\mgl$ upto 1950 GeV subject to the lightest neutralino mass close to 1800 GeV 
\cite{ATLASCollaboration:2016wlb,Khachatryan:2016ojf,ATLAS:2016fks}, which is a significant 
improvement on the bounds obtained after the 8 TeV run with $20.3~{\rm fb^{-1}}$ integrated luminosity 
\cite{Aad:2015hea,Khachatryan:2016hns}. We note that, 
in order to derive the limits from the collider data, the experimental collaboration considers signal events coming from 
gluino pair production only, while assuming the rest of the coloured sparticles {\sl viz.} squarks to be much heavier to contribute 
to the signal. 
The robustness of the signal however does not differentiate whether such a heavy SUSY spectrum (leaving aside the gravitino) 
are closely spaced in mass or have a widely split mass spectrum, and whether it is just a single sparticle state that contributes 
to the signal or otherwise. We intend to impress through this work that such a signal would also be able to distinguish such 
alternate possibilities quite efficiently.

In an earlier work while assuming a similar compression in the sparticle spectrum \cite{Dutta:2015exw} we had shown 
that in order to get a truly compressed\footnote{Mass gap between the heaviest coloured 
sparticle and the LSP neutralino has to be around 100 GeV.}  pMSSM spectrum consistent with a 125 GeV Higgs boson 
and the flavour and dark matter (DM) constraints, one has to have the $\lspone$ mass at or above 2 TeV with the entire coloured 
sector lying slightly above. Such a spectrum is now seemingly of interest given the present experimental bounds obtained in 
$\grav$ LSP scenario\footnote{Note that the bounds on the squark-gluino masses in the compressed region with 
$\lspone$ LSP are still much weaker. In such cases, the gluinos and first two generation squarks are excluded upto 650 GeV 
and 450 GeV respectively \cite{Aaboud:2016zdn}.}. 
In this work, we aim to extend our previous study by adding to the spectrum, a $\grav$ LSP with mass, at most, in the eV-keV range. 
Rest of the pMSSM spectrum lies above the TeV range to be consistent with the experimental bounds. This is in 
contrast to existing studies done earlier for gravitino LSP which we compare by studying the prospects of uncompressed spectra 
having relatively larger mass gaps between the coloured sparticles 
and $\lspone$, but with event rates similar to that of the compressed spectra. Since the kinematics of 
the decay products in the two cases are expected to be significantly different, we 
present some kinematical variables which clearly distinguish a compressed spectrum from an uncompressed one, 
in spite of comparable signal rates in both cases. 

The paper is organised in the following way. In Section~\ref{sec:grvlsp} we discuss about the phenomenological aspects of a 
SUSY spectrum with gravitino LSP and then move on to study the variation of the branching 
ratios of squark, gluino and the lightest neutralino into gravitino associated and other relevant decay modes. In 
Section~\ref{sec:bp} we present some sample benchmark points representative of our region of interest consisting of both 
compressed and uncompressed spectra that are consistent with the existing constraints. Subsequently, in Section~\ref{sec:coll} 
we proceed to our collider analysis with these benchmark points and present the details of our simulation and obtained results. 
Finally, in Section~\ref{sec:sumcon} we summarise our results and conclude.
%%%%%%%%%%%%%%%%%%%%%%%%%%%%%%%%%%%%%%%%%%%%%%%%%%
\section{Compressed spectrum with a gravitino LSP}
\label{sec:grvlsp}
%%%%%%%%%%%%%%%%%%%%%%%%%%%%%%%%%%%%%%%%%%%%%%%%%%
The NLSP decaying into a gravitino and jets/leptons/photons give rise to very distinct signals at the LHC. Both the ATLAS 
and CMS collaborations have studied these signal regions for a hint of GMSB-like scenarios \cite{Aad:2012zza,Aad:2015hea,
CMS:2016igk,Khachatryan:2016hns,CMS:2016nij,ATLASCollaboration:2016wlb,Khachatryan:2016ojf,ATLAS:2016fks}. 
Note that, a pure GMSB like scenario is now under tension after the discovery of the 125 GeV Higgs boson \cite{Arbey:2011ab,
Ajaib:2012vc,Djouadi:2013jqa}. It is very difficult to fit a light Higgs boson 
within this minimal framework, mostly because of small mixing in the scalar sector. As a consequence, 
the stop masses need to be pushed to several TeV in order to obtain the correct Higgs mass, thus rendering such scenarios 
uninteresting in the context of LHC. However, some variations of the pure GMSB scenario are capable of solving the Higgs 
mass issue and can still give visible signals within the LHC energy range \cite{Albaid:2012qk,Byakti:2013ti}. 
Since we are only interested in the phenomenology of these models here, a detailed discussion on their
theoretical aspects is beyond the scope of this paper. 
   
Although the lightest neutralino ($\lspone$) is the more popular DM candidate in SUSY theories, gravitino ($\grav$) as the LSP  
has its own distinct phenomenology. The $\grav$ is directly related to the effect of SUSY breaking via gauge mediation
and all its couplings are inversely proportional to the Planck mass ($\sim 10^{18}$ GeV) and thus considerably suppressed.
The hierarchy of the sparticle masses depend on the SUSY breaking mechanism and can result in $\grav$ 
getting mass which is heavier, comparable or lighter than the other superpartners. Thus if it happens to be the LSP 
in the theory, $\grav$ can also be a good DM candidate \cite{Baltz:2001rq,Viel:2005qj,Covi:2009bk,Covi:2010au,Arvey:2015nra,LCovi} 
making such scenarios of considerable interest in the context of the LHC. 
In addition, having $\grav$ as a DM candidate also relaxes the DM constraints on the rest of the SUSY spectrum by a great deal, 
allowing them to be very heavy while being consistent with a light $\grav$ DM. However, a very light $\grav$ is mostly considered to be 
warm DM. Present cosmological observations require a light gravitino to have a mass close to a few keV 
\cite{Boyarsky:2008xj,Baur:2015jsy} at least, 
if it has to explain the cold DM relic density. However, the kinematic characteristics of events when the NLSP decays into a 
gravitino are mostly independent of whether the gravitino is in the keV range or even lower in mass. Some special situations 
where the difference is of some consequence have been discussed in Section~\ref{sec:evgr}. Of course, the presence of a gravitino 
much lighter than a keV will require the presence of some additional cold DM candidate.

Note that with $\grav$ as the LSP  decay branching ratios (BR) of the sparticles can be significantly modified since they 
can now decay directly into $\grav$ instead of decaying into $\lspone$, which may significantly alter their collider signals.
The decay width ($\Gamma$) of a sparticle, scalar($\widetilde f$) or gaugino($\widetilde V$), decaying into their respective 
SM counterparts, chiral fermion($f$) or gauge boson($V$), and  $\grav$ is given by \cite{Drees:2004jm}
\begin{eqnarray}
\Gamma(\widetilde f\to f \grav) = \frac{1}{48\pi}\frac{m^5_{\widetilde f}}{M^2_{Pl}m^2_{\grav}}\left[1-\left(\frac{m_{\grav}}{m_{\widetilde f}}\right)^2 \right ]^2 \\
\Gamma(\widetilde V\to V \grav) = \frac{1}{48\pi}\frac{m^5_{\widetilde V}}{M^2_{Pl}m^2_{\grav}}\left[1-\left(\frac{m_{\grav}}{m_{\widetilde V}}\right)^2 \right ]^3
\end{eqnarray}
where $M_{Pl}$ is the Planck scale. 
Thus it is evident that this decay mode starts to dominate once the sparticles become very heavy and the $\grav$ becomes 
light.   
%%%%%%%%%%%%%%%%%%%%%%%%%%%%%%%%%%%%%%%%%%%%%%%%%%
\subsection{Relevant Branching Ratios}
\label{sec:spbr}
%%%%%%%%%%%%%%%%%%%%%%%%%%%%%%%%%%%%%%%%%%%%%%%%%%
In this section, we discuss the variation of the branching ratios (BR) of various sparticles into the LSP gravitino. 
Since in this analysis we aim to study the production of the coloured sparticles and their subsequent decays into 
the $\grav$ via $\lspone$, the decay modes of $\gl$, $\sq$ and $\lspone$ are of our primary interest. While 
considering the decay modes, we focus on a simplified assumption that the decaying coloured sparticle is the 
next-to-next-lightest supersymmetric particle (NNLSP) with $\lspone$ as the NLSP and $\grav$ as the LSP. 
The BR computation and spectrum generation was done using {\tt SPheno} \cite{Porod:2002wz,
Porod:2003um,Porod:2011nf} for a phenomenological MSSM (pMSSM) like scenario with one 
additional parameter, i.e, the gravitino mass ($\mgrav$). 
%%%%%%%%%%%%%%%%%%%%%%%%%%%%%%%%
\subsubsection{Variation of BR($\gl\to g\grav$)}
\label{sec:glbr}
%%%%%%%%%%%%%%%%%%%%%%%%%%%%%%%%
\begin{figure}[h!]
\begin{center}
\includegraphics[scale=0.55]{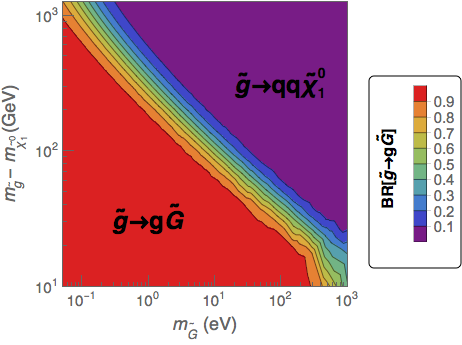}
\caption{Variation of BR($\gl\to g\grav$) and BR($\gl\to q\bar q\lspone$) shown colour coded in 
$\Delta m_{\gl\lspone}$ - $\mgrav$ plane.}
\label{fig:gl_br}
\end{center}
\end{figure}
%%%%%%%%%%%%%%%%%%%%%%%%%%%%%%%%
In Fig.~\ref{fig:gl_br} we show the variation of two relevant gluino decay mode channels {\it viz.} $\gl\to g\grav$ and 
$\gl\to q\bar q\lspone$  where all the squarks are heavier, as a function of $\Delta m_{\gl\lspone}=\mgl - \mlspone$ 
and $\mgrav$. 
The gluino mass has been fixed to $\mgl$=2500 GeV while $\mlspone$ has been varied such that 
$\Delta m_{\gl\lspone}$ varies within 10-1500 GeV. Note that 
%like the squarks, all the other neutralino and chargino states are also assumed to be heavier than the gluino and 
the $\lspone$ is considered to be dominantly bino-like.  
In the absence of its two-body decay mode into squark-quark pairs, the gluino can only decay via $\gl\to g\grav$ or 
$\gl\to q\bar q\lspone$. The other two-body decay mode $\gl\to g\lspone$ being loop suppressed, remains mostly subdominant 
compared to these two decay modes. Hence, only the two relevant channels are shown 
in the figure. Note that, BR($\gl\to q\bar q\lspone$) includes the sum of all the 
off-shell contributions obtained from the first two generation squarks which in this case lie about 100 GeV above $\mgl$. As the gravitino mass gets heavier, 
BR($\gl\to g\grav$) decreases since, the corresponding partial width is proportional to 
the inverse square of  $m_{\grav}$. Similarly, as $\mlspone$ keeps increasing, 
BR($\gl\to q\bar q\lspone$) goes on decreasing. Note that, the BR for the 3-body decay 
mode can decrease further with increase in the corresponding squark masses. However, 
even for a keV $\grav$, BR($\gl\to g\grav$)  can remain significantly large provided 
there is sufficient compression in the mass gap ($\Delta m_{\gl\lspone}\sim$10 GeV) 
as seen in Fig.~\ref{fig:gl_br}.  
%%%%%%%%%%%%%%%%%%%%%%%%%%%%%%%%
\subsubsection{Variation of BR($\sq_{L/R}\to q\grav$)}
\label{sec:sqbr}
%%%%%%%%%%%%%%%%%%%%%%%%%%%%%%%%
\begin{figure}[h!]
\begin{center}
\includegraphics[scale=0.45]{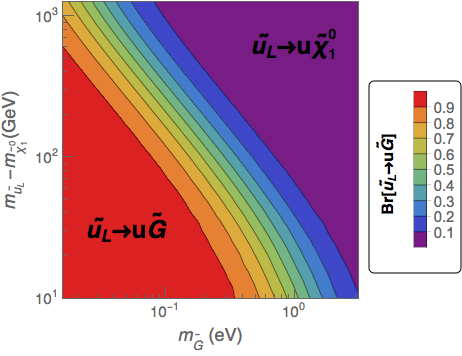} 
\includegraphics[scale=0.45]{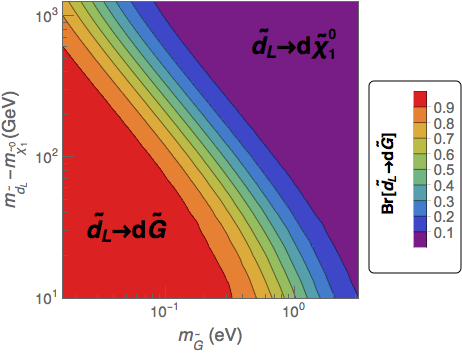}  
\includegraphics[scale=0.45]{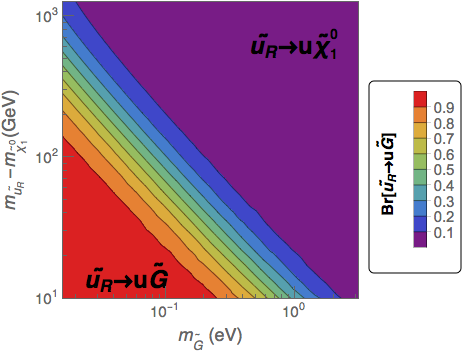} 
\includegraphics[scale=0.45]{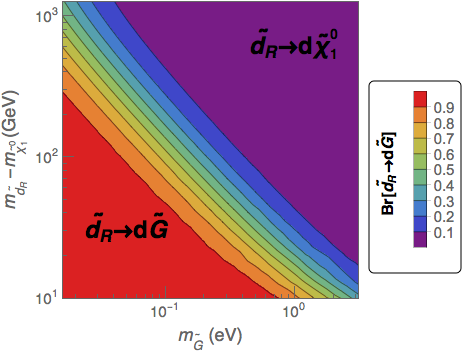}
\caption{Variation of BR($\sq_{L/R}\to q\grav$) and BR($\sq_{L/R}\to q\lspone$)  
in the plane $\Delta m_{\sq\lspone}$ - $\mgrav$. 
The plots on the left show the distributions corresponding to the up-squarks and the plots on the right show the same 
for the down-squarks.}
\label{fig:sq_br}
\end{center}
\end{figure}
%%%%%%%%%%%%%%%%%%%%%%%%%%%%%%%
Next we look into the relevant decay modes of the first two generation squarks\footnote{Since the production cross-section of the 
third generation squarks are substantially smaller than those of the first two generations, 
we do not consider the production of the stop and sbottom states. Hence we only discuss the decays of $\tilde{u}_{L/R}$ and $\tilde{d}_{L/R}$.} when they are 
the NNLSP's. In this case, we assume that the gluino is heavier than the squarks, so that the 
dominant two-body decay modes available to the squarks are $\sq_{L/R}\to q\grav$ and $\sq_{L/R}\to q\lspone$. 
Unlike the previous case, here the gravitino decay branching ratio has competition from another two-body decay mode. 
Although the decay into $\grav$ does not depend on the L and R-type of the squarks, BR($\sq_{L/R}\to q\lspone$) is expected 
to be different depending on the composition of the $\lspone$. For simplicity, we choose the $\lspone$ to be purely bino-like
as before. The squark masses are fixed at $m_{\sq} = 2500$ GeV and the NLSP mass, $\mlspone$ is varied as before 
such that $\Delta m_{\sq\lspone}=\msq - \mlspone$ varies in a wide range, 10-1500 GeV.  
The branching probabilities are shown in Fig.~\ref{fig:sq_br} where the plots on the left (right)   
shows the decay branching ratios of $u_{L/R} \,\, (d_{L/R})$.    
As the coupling of $\sq_L$ with the SM-quark and bino-component of $\lspone$ is proportional to 
$\sqrt{2}g~{\rm tan}\theta_W(I_{3q}-e_q)$ while that of $\sq_R$ is proportional to $\sqrt{2}g~{\rm tan}\theta_W e_q$, 
where $g,e_q~{\rm and}~I_{3q}$ represents SU(2) gauge coupling, electric charge of the SM-quark and its isospin respectively \cite{Drees:2004jm}, we find a noticeable variation
in decay probabilities of $\sq_L$ and $\sq_R$ for the same choice of mass spectrum.
This implies that the right-handed squarks couple more strongly with the $\lspone$ compared to the left-handed ones. 
As a result, although the partial decay widths of the squarks decaying into gravitino and quarks are identical for 
squarks of similar mass, the corresponding BR vary slightly depending on their handedness. 
This feature is evident in Fig.~\ref{fig:sq_br}. 
The coupling strength of $\widetilde u_R$ with $\lspone$ is larger by a factor of four compared to that of $\widetilde u_L$. The same coupling 
corresponding to $\widetilde d_R$ is larger by a factor of two compared to that of $\widetilde d_L$. 
Hence the difference in the BR distributions is more manifest for the up-type squarks. The magnitude of the coupling 
strengths corresponding to $\widetilde u_L$ and $\widetilde d_L$ are exactly same and hence we have obtained similar distributions 
corresponding to those. 

The BR distributions indicate that as we go on compressing the SUSY spectrum, the gravitino decay mode becomes 
more and more relevant but only if its mass is around or below the eV range. 
We, therefore, conclude that for a keV $\grav$, the decay mode $\gl\to g\grav$ may be of importance 
but only for the cases where the gluino mass lies very close to the NLSP neutralino mass. For the 
first two generation squarks and a keV $\grav$, the BR($\sq_{L/R}\to q\grav$) is very small and the decay of the 
squarks into $\lspone$ dominates in the absence of a lighter gluino. As evident, the gravitino decay mode can be 
of significance for LHC studies if $\mgrav\sim$ eV. However, such a light $\grav$ is strongly disfavoured from DM 
constraints as mentioned before. 
%%%%%%%%%%%%%%%%%%%%%%%%%%%%%%%%
\subsubsection{Variation of BR($\lspone\to X\grav$)}
\label{sec:ntbr}
\begin{figure}[t!]
\includegraphics[width=4.9cm, height=5cm]{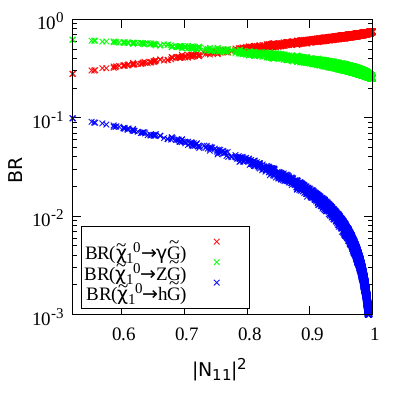}  
\includegraphics[width=4.9cm, height=5cm]{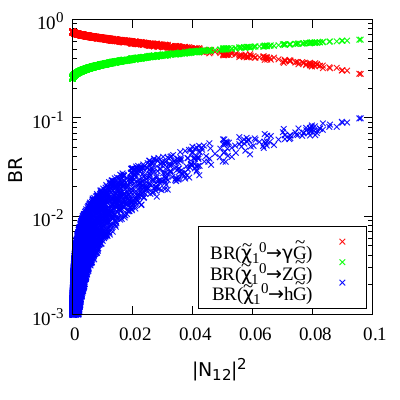}
\includegraphics[width=4.9cm, height=5cm]{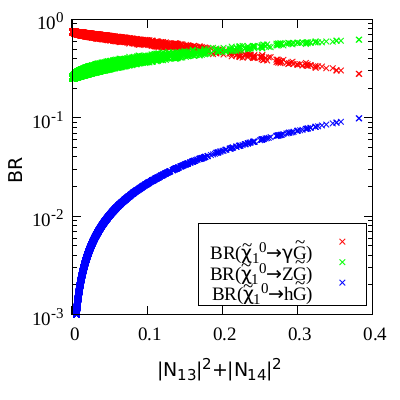}
\caption{Vatiation of the three relevant BRs of $\lspone$ decay modes with its 
bino, wino and higgsino components. The red, green and blue lines correspond to
BR($\lspone\to\gamma\grav$), BR($\lspone\to Z\grav$) and BR($\lspone\to h\grav$) 
respectively.}
\label{fig:neut_br}
\end{figure}

%%%%%%%%%%%%%%%%%%%%%%%%%%%%%%%%
The last two subsections point out the situations where the NLSP can be bypassed in the decay of strongly interacting 
superparticles. Such events tend to reduce the multiplicity of hard photons in SUSY-driven final states. In contrast, in the case where the 
SUSY cascades lead to a $\lspone$ NLSP,
the $\lspone$ may further decay into gravitino along with a $Z$, $\gamma$ or the Higgs boson ($h$) depending upon its 
composition\footnote{In principle, $\lspone$ may decay into the other neutral Higgs states also which we assume to be heavier.}. 
The $h$-associated decay width is entirely dependent on the higgsino component of $\lspone$ while $\Gamma(\lspone\to \gamma\grav)$ depends entirely on the bino and wino component of $\lspone$ whereas the $Z$-associated decay 
width has a partial dependence on all the components that make up the $\lspone$. 
The functional dependence on the different composition strengths of $\lspone$ in 
its decay width can be summarised as \cite{Drees:2004jm}:  
\begin{eqnarray}
\Gamma(\lspone\to \gamma\grav)&&\propto |N_{11}cos\theta_W + N_{12}sin\theta_W|^2  \\
\Gamma(\lspone\to Z\grav) &&\propto \left (|N_{11}sin\theta_W - N_{12}cos\theta_W|^2 + \frac{1}{2}|N_{14}cos\beta - 
N_{13}sin\beta|^2\right ) \\
\Gamma(\lspone\to h\grav) &&\propto |N_{14}sin\alpha - N_{13}cos\alpha|^2 
\end{eqnarray}
where, $N_{ij}$ are the elements of the neutralino mixing matrix, $\theta_W$ is the Weinberg mixing angle, $\alpha$ is the neutral Higgs mixing angle 
and $\beta$ corresponds to the ratio of the up and down type Higgs vacuum expectation values (VEVs).  
Note that the partial decay widths are proportional to ${m^5_{\lspone}}/{(M^2_{Pl}m^2_{\grav})}$ %$\frac{m^5_{\lspone}}{M^2_{Pl}m^2_{\grav}}$ 
and hence if $\mgrav$ is too large, the total decay width of $\lspone$ may become too small such that it will not decay within 
the detector. Although the decay width is also dependent upon $\mlspone$, one finds that for a 2500 GeV $\lspone$, and a MeV 
$\grav$ the neutralino becomes long-lived. In Fig.~\ref{fig:neut_br} we show the variation of the three relevant BRs with the composition of the $\lspone$. 
%%%%%%%%%%%%%%%%%%%%%%%%%%%%%%%%
%%%%%%%%%%%%%%%%%%%%%%%%%%%%%%%%
Here we have varied $M_1, \, M_2$ and $\mu$ in the range $\left[2 : 2.5\right]$ TeV with the condition 
$\mu > M_2 > M_1$ such that $\lspone$ is bino-like most of the time with different 
admixtures of wino and higgsino components. The other relevant mixing 
parameter ${\rm tan}\beta$ is kept fixed at 10. The red, green and blue colours 
correspond to BR($\lspone\to\gamma\grav$), BR($\lspone\to Z\grav$) and 
BR($\lspone\to h\grav$) respectively. $|N_{11}|^2$ indicates the bino-fraction in the 
composition of $\lspone$. 
Similarly, $|N_{12}|^2$ and $|N_{13}^2|+|N_{14}|^2$ represent the wino and higgsino 
components respectively. 
As can be clearly seen from the plots, obtaining 100\% BR($\lspone\to\gamma\grav$) 
is not possible even if the bino and(or) wino components are close to 1, since 
the $Z$-mode is always present. However, the $h$-associated decay channel can be easily 
suppressed with a relatively larger $\mu$. Motivated by this behaviour of the BRs, we 
choose to work with a signal consisting of at least one photon for our collider analysis. 
In our case, the $\lspone$ being dominantly bino-like, it decays mostly into a 
$\gamma$ and a $\grav$. However, the $Z\grav$ decay mode has a substantial 
BR ($\sim 25\%$). The higgsino admixture in $\lspone$ being small, 
the $h\grav$ decay mode is not considered in this work. However it is worth noting that this particular channel can be the 
dominant mode for a higgsino-dominated NLSP and could also be an interesting mode 
of study, which we leave for future work.

%%%%%%%%%%%%%%%%%%%%%%%%%%%%%%%%
\section{Benchmark Points}
\label{sec:bp}
%%%%%%%%%%%%%%%%%%%%%%%%%%%%%%%%
For our analysis we choose a few benchmark points that would represent the salient 
features of a compressed sparticle spectrum with varying compression strengths while also 
categorically defining a few points that are more in line with current SUSY searches 
with $\grav$ LSP by the CMS and ATLAS collaborations at the LHC. We insure that our 
benchmark choices are consistent with all existing experimental constraints. 
We consider both compressed and uncompressed spectra, with bino-like $\lspone$ as the NLSP and a keV gravitino as the
LSP and warm dark matter candidate. For one of the benchmarks, we also show the effect of an 
eV mass gravitino LSP. The final benchmarks used in this study are shown in Table~\ref{tab:bp_comp}.
 
The mass spectrum and decays of the sparticles are computed using {\tt SPheno}-v3.3.6 
\cite{Porod:2002wz,Porod:2003um,Porod:2011nf}. We restrict the light CP-even Higgs 
mass to be in the range 122-128 GeV, i.e, within 3-$\sigma$ range of the measured 
Higgs mass \cite{Aad:2012tfa,Chatrchyan:2012xdj,Aad:2015zhl,ATLAS-CONF-2016-081} 
and including theoretical uncertainty of $\sim$ 4 GeV. 
Note that when the mass spectrum is compressed, all squark$/$gluino (which are nearly 
degenerate in mass) production channels contribute significantly to the signal. 
For all the benchmark points, the squarks and gluino decay directly or via cascades to the bino-like $\lspone$ NLSP. 
The $\lspone$ then dominantly decays to a photon and gravitino and, to a lesser extent, a $Z$ boson and 
gravitino. This leads to either a mono-photon or a diphoton signal with jets and $\met$ which defines our signal.
To evade constraints from photon(s) searches at the LHC for simplified models 
\cite{Aad:2012zza,Aad:2015hea,CMS:2016igk,Khachatryan:2016hns,CMS:2016nij,ATLASCollaboration:2016wlb,Khachatryan:2016ojf,
ATLAS:2016fks}, we require the sparticles in a compressed spectrum such as ours, 
to be much heavier than the existing experimental limits. We have checked this for 
our spectra represented by the benchmark points, with the NLSP mass lying in the 
range $2.4 - 2.6$ TeV with varied masses and hierarchy of the coloured sparticles with 
respect to the NLSP.
Amongst them, {\bf C6} is the utmost compressed spectra, with a mass gap, $\Delta M_i$ $\sim$ 6 GeV 
between the coloured sparticles and the NLSP of mass 2462 GeV, 
followed by {\bf C2}, {\bf C5} where the mass gap is in the range of 40-50 GeV and the NLSP masses are 2428 and 2526 GeV 
respectively. We have also considered benchmarks {\bf C1}, {\bf C3} and {\bf C4} such that the mass gap between 
the coloured sparticles and NLSP are slightly higher and lie in the range of 100-200 GeV. 

We also choose 
various possible mass hierarchical arrangements of the squarks and gluino to accommodate different cascades contributing to the signal.
For example, {\bf C1} and {\bf C3} have different squark-gluino mass hierarchical stuctures in the strong sector. 
This leads to different jet distributions in the two cases. {\bf C2} and {\bf C5}, on the other hand, are similar in the 
arrangement of the sparticles, however placed within 50 GeV from the NLSP, which represents a much more compressed scenario.
Finally we consider two uncompressed spectra {\bf U1}, {\bf U2} with NLSP mass $\sim$ 700 GeV and $\sim$ 1200 GeV and 
gluinos with mass $\sim$1.4 TeV and $\sim$1 TeV above the NLSP respectively.  
Since the photons arise from the NLSP decays, a heavier NLSP gives rise to a harder photon, having better chances of 
passing the analysis cuts. Thus the difference in the signal cross-sections differ on account of 
the difference in hardness of the photons and the resulting cut efficiencies in these two cases.

Benchmark points {\bf U1}, {\bf U2} are in fact
replications of the simplified scenarios that are considered by experimental collaborations to put limits on SUSY particle masses. 
For both these benchmark points, we have kept the squarks very heavy ($\sim 4-5$ TeV) so that the gluino pair production is the only dominant
contributing channel. However, we have only focussed on uncompressed spectra with event rates comparable 
to those of the compressed spectra. Since the large mass gap between the gluino and NLSP allow for multiple hard jets to be 
produced as opposed to the compressed case, we further exploit this feature to differentiate compressed from uncompressed scenarios with comparable event rates during signal analysis.
%%%%%%%%%%%%%%%%%%%%%%%%%%%%%%%%
\begin{center}
\begin{table}[ht]
\small
\begin{tabular}{|c|c|c|c|c|c|c|c|c|c|c|} 
\cline{2-9}
\multicolumn{1}{c|}{} &
\multicolumn{6}{|c|}{\bf{Compressed spectra}} & 
\multicolumn{2}{|c|}{\bf{Uncompressed spectra}} \\
\hline
\bf{Parameters}  & \bf{C1} & \bf{C2} & \bf{C3} & \bf{C4} & \bf{C5} & \bf{C6} & \parbox[c]{1.9cm}{\bf{\hspace{0.7cm}U1}}  & \bf{U2}  \\ 
\hline
$M_1$  & 2623 & 2451 & 2671 &2608& 2550 & 2486 &704 & 1200  \\
$M_2$  & 2710 & 2610 & 2710 & 2710 &2610 & 2610&  2310 & 2310 \\    
$M_3$  & 2480 & 2280& 2560 & 2601 &2380 &2285 &1747 & 1747\\
$A_t$  & 2895& 2895& -3295   & -3750& -3197 &-2895 & 2895 & 2895\\
$\mu$  & 4000 & 4000& 4000  & 4000 &3500 & 4000 &3000&  3000\\ 
$\tan\beta$  & 15 & 15  & 9 &  6  &25 &15 &15 & 15\\
$M_A$ & 2500& 2500& 1800 &  1800 &2500  &2500 &2500 &  2500\\
\hline
\newline
$m_{\widetilde g}$  & 2678 & 2456& 2746 & 2783& 2562& 2468 &2102 & 2102 \\
$m_{\widetilde q_{L}}$  &2729 & 2468& 2734 & 2753 & 2571 & 2467 &4721 & 4721 \\
$m_{\widetilde q_{R}}$  &2727& 2466& 2730  & 2751 &2574 & 2468 &4742 & 4742\\
$m_{\widetilde t_1}$  &2707 &2457 & 2652 & 2625  & 2532 & 2543 &4680 & 4678 \\
$m_{\widetilde t_2}$  & 2837& 2593&  2857 & 2863& 2718 & 2725 &4767 & 4765 \\
$m_{\widetilde b_1}$  & 2787& 2501 & 2782  & 2778&2594 & 2598 &4560 & 4558 \\
$m_{\widetilde b_2}$  & 2846&2570&  2846 & 2846& 2677& 2669 & 4746& 4744 \\
$m_{\widetilde\ell_{L}}$  &2703  &2452   & 2703 & 2703&2572 & 2503 & 4335& 4336\\
$m_{\widetilde\ell_{R}}$  & 2700 &2455 & 2700  & 2700&2585 & 2495 &4365 &  4366\\
$m_{\widetilde \tau_1}$ & 2706 & 2443&  2707 & 2709 &2600 & 2576 &4332 & 4332 \\
$m_{\widetilde \tau_2}$  &2882 &2514 & 2882 & 2881 &2671 & 2622 &4375 & 4375\\
$m_{\widetilde\nu_{L}}$  & 2701 & 2450&  2701 & 2701 &2570 & 2501 &4335& 4335 \\
$m_{\widetilde\chi^0_1}$& 2600 &2428 & 2646   &2585&2526 &2462 &699& 1191\\
$m_{\widetilde\chi^0_2}$ & 2726 &  2614  & 2724& 2724&2619 & 2617& 2383& 2383\\
$m_{\widetilde\chi^{\pm}_1}$  & 2726 & 2614  & 2725 &2724& 2619 & 2617& 2382 & 2382\\
\hline
$m_h$ & 123&123 & 124& 124&125&124 &125 & 125\\
$\Delta M_i $  & 129& 40 &100&198& 48 &6 &1403& 911\\
\hline
\end{tabular}
\caption{Low energy input parameters and the relevant sparticle masses, (in GeV), for the compressed (\textbf{C$_{i}$}, i = 1,...,6) 
and uncompressed (\textbf{U1}, \textbf{U2}) benchmarks. Here, $\Delta M_i = m_i-\mlspone$ %and  $\Delta M_j = m_j-\mlspone$,   
where $m_i$ represents the mass of the heaviest coloured sparticle $(\widetilde{g} / 
\widetilde{q}_{k}, \text{(k = 1,2)})$ and $\mlspone$, the mass of the NLSP. For all benchmarks, the gravitino mass, $m_{\grav}$ = 1 keV.}
\label{tab:bp_comp}
\end{table} 
\end{center}
%%%%%%%%%%%%%%%%%%%%%%%%%%%%%%%% 
%%%%%%%%%%%%%%%%%%%%%%%%%%%%%%%%%%%%%%%%%%%%%%%%%%
\section{Collider Analysis}
\label{sec:coll}
%%%%%%%%%%%%%%%%%%%%%%%%%%%%%%%%%%%%%%%%%%%%%%%%%%
We look for multi-jet signals associated with very hard photon(s) and 
missing transverse energy ($\met$) in the context of SUSY with gravitino as the LSP. 
For such GMSB kind of models with a keV gravitino, a very clear signature arises from 
the decay of the NLSP neutralino into a photon and a gravitino. If the NLSP-LSP mass difference is large enough, two hard photons would 
appear in the final state at the end of a SUSY cascade. The lightest neutralino, if bino-like, decays 
dominantly into a photon and gravitino ($\sim$75$\%$) while a small fraction decays into 
$Z$ boson and gravitino ($\sim25\%$). For cases with $\lspone$ having a significant higgsino component, we get
comparable branching fractions for its decay into $Z$ boson or a Higgs boson,  
besides photons, along with $\grav$. For simplicity, we have considered a bino-like 
$\lspone$ as the NLSP. 
Note that the signal strength consisting of very hard photons in the final state can be affected by the composition of the NLSP as we have discussed
before. The $\lspone$ decay into a $Z\, \grav$ however still remains relevant for the bino-like $\lspone$ and as a result, gives 
rise to a monophoton signal at the LHC along with the diphoton channel, associated with large missing transverse energy. 
The existing LHC constraints in such scenarios have already pushed the $\lspone$-$\sq$-$\gl$ mass bounds above 1.5 TeV 
which automatically result in a large $\lspone$ - $\grav$ mass gap. This gives rise to very high $p_T$ photons 
in the final states, which are very easy to detect and also highly effective to suppress the SM background events.  

In this work, we consider six benchmark points for compressed spectra (\textbf{C1 - C6}) such that 
the entire coloured sector (apart from $\hstop$ and $\hsbot$) lie within 200 GeV of the $\lspone$ (m$_{\chi_1^0} \sim$ 2.4 - 2.6 TeV). 
We then estimate signal rates of final state events with at least one or more hard photons arising from 
all possible squark-gluino pair production modes. We also study a couple of uncompressed spectra (\textbf{U1,U2}) 
such that both the compressed and uncompressed spectra produce similar event rates for our signal.
In these spectra, the NLSP mass is around 700 and 1200 GeV respectively and the gluino is the lightest 
coloured sparticle having a large ($\sim$ 1-1.4 TeV) mass gap with the NLSP. The squarks are chosen to be 
heavier (4-5 TeV) and are essentially decoupled from rest of the spectrum. 
The large mass gap between the NLSP and the coloured sector ensures multiple hard jets from their
decay cascades besides the hard photons. Thus with different mass gaps and squark-gluino hierarchy 
among the compressed and uncompressed spectra, the jet profiles are expected to be significantly different for the benchmark points. 
Following the existing ATLAS analysis \cite{ATLAS:2016fks}, which provides the most stringent constraint on the SUSY spectrum 
with a light gravitino LSP, we determine the signal event rates for our choice of benchmark points. 
Since we have also chosen compressed and uncompressed spectra such that the final state event rates are equal
or comparable after analysis, it is {\it a priori} difficult to determine which scenario
such a signal reflects. Keeping this in mind, we propose a set of kinematic
variables, besides the usual kinematic ones like $\met$ and $M_{Eff}$, which highlight the distinctive features of compression in a 
SUSY spectra over an uncompressed one with $\grav$ as the LSP, although both have comparable signal rates.

%%%%%%%%%%%%%%%%%%%%%%%%%%%%%%%%%%%%
\subsection{Simulation set up and Analysis}
\label{sec:simul}
%%%%%%%%%%%%%%%%%%%%%%%%%%%%%%%%%%%%
We consider the pair production and associated production processes of all coloured sparticles at $\sqrt{s}=13$ TeV LHC. Parton level events 
are generated using {\tt Madgraph5} (v2.2.3) \cite{Alwall:2011uj,Alwall:2014hca} for the following 
processes with upto two extra partons at the matrix element level:
\begin{equation*}
 p \, p \rightarrow \widetilde{q}^{ *}\,{\widetilde {q}}, \,\, \widetilde{q }\,\widetilde{g},\,\,  \widetilde{q }\,\widetilde{{q}},\,\,
 \widetilde{q}^{ *}\,\widetilde{q}^{ *}, \,\, {\widetilde {q}}^{ *}\,\widetilde{g},\,\,  \widetilde{g }\,\widetilde{g}
\end{equation*}
We reject any intermediate resonances at the matrix element level, which may arise in the decay cascades of the sparticles
from two or more different processes, to avoid double counting of Feynman diagrams to the processes.
The parton level events are then showered using {\tt Pythia} (v6) \cite{Sjostrand:2006za}. To correctly 
model the hard ISR jets and reduce double counting of jets coming from the showers as well as the matrix element 
partons, MLM matching \cite{Mangano:2006rw,Hoche:2006ph} of the shower jets and the matrix element jets have been 
performed using the shower-k$_{T}$ 
algorithm with p$_{T}$ ordered showers by choosing a matching scale (QCUT) 120 GeV \cite{madgraph_matching}. 
The default dynamic factorisation 
and renormalization scales \cite{madgraph_scale} have been used in Madgraph whereas the PDF chosen is CTEQ6L \cite{Pumplin:2002vw}.
After the showering, hadronisation and fragmentation effects performed by Pythia, subsequent detector simulation 
of the hadron level events are carried out by the fast simulator {\tt Delphes}-v3.3.3 
\cite{deFavereau:2013fsa,Selvaggi:2014mya,Mertens:2015kba}.  The jets are reconstructed 
using {\tt Fastjet} \cite{Cacciari:2011ma} with a minimum $p_{T}$ of 20 GeV in a cone of $\Delta R = 0.4$ using the 
{\sl anti}-$k_t$ algorithm \cite{Cacciari:2008gp}. 
The charged leptons ($e,\mu$) are reconstructed in a cone of $\Delta R = 0.2$ with the maximum amount of energy 
deposit allowed in the cone limited to 10\% of the $p_{T}$ of the lepton. Photons are reconstructed in a cone of 
$\Delta R = 0.4$, with the maximum energy deposit in the cone as per ATLAS selection criteria \cite{ATLAS:2016fks}.

For background estimation, we focus on the most dominant SM backgrounds for photon(s) + jets + $\met$ signal at 13 TeV LHC, 
such as: $\gamma +$ $\leq$ 4 jets, $\gamma \gamma + $ $ \leq$ 3 jets, $W \gamma +$ $\leq$ 3 jets, 
$Z \gamma + $ $\leq$ 3 jets and $t \bar{t} \gamma$ $+$ jets. The sort of extremely hard $p_{T}$ photons that we expect 
in our signal events, are unlikely to be present in SM processes in abundance and the hard photons will arise 
mostly from the tails of the $p_{T}^{\gamma}$ distributions. Hence in order to obtain a statistically exhaustive event 
sample, we choose a hard $p_{T_{\gamma}} > 200$ GeV cut as a preselection for the parton level events for the leading 
photon while generating the background events. For MLM matching of the jets, the matching scale was chosen in the 
range 30-50 GeV as applicable for electroweak SM processes. 

Some other SM processes, such as QCD, $t\bar{t}+$jets, $W+$jets, $Z+$jets, in spite of having no direct sources of hard 
photons, may also contribute to the background owing to their large production cross-sections coupled with  
mistagging of jets or leptons leading to fake photons. However, the cumulative effect of hard $p_T^{\gamma}$ 
as well as $\met$ and $M_{Eff}$ requirement renders these contributions negligible. 
%%%%%%%%%%%%%%%%%%%%%%%%%%%%%%%% 
\subsubsection*{Primary Event selection criteria}
\label{sec:selcn}
%%%%%%%%%%%%%%%%%%%%%%%%%%%%%%%% 
We identify the charged leptons ($e,\mu$), photons and jets as per the following selection criteria ({\bf A0})
for signal and background events alike: 
\begin{itemize}
\item Leptons ($\ell = e,\mu$) are selected with $p_T^{\ell} > 25$  GeV, $|\eta^{e}| < 2.37$ and $|\eta^{\mu}| < 2.70$ and excluding 
the transitional pseudorapidity  window $1.37 < |\eta^{\ell}| < 1.52$ between the ECAL barrel and end cap of the calorimeter.
\item Photons are identified with $p_T^{\gamma} > 75$ GeV and $|\eta^{\gamma}| <  2.47$ excluding $1.37 < |\eta^{\gamma}| < 
1.52$.
\item  Reconstructed jets have $p_T^{j} > 30$ GeV and lie within $|\eta^j| < 2.5$.
\item All reconstructed jets have a large azimuthal separation with $\vec{\slashed{E}}_T$, given by 
$\Delta\phi(\vec{{\rm jet}},\vec{\slashed{E}}_T)  > 0.4$ to reduce fake contributions to missing transverse energy 
arising from hadronic energy mismeasurements.
\item The jets are separated from other jets by $\Delta R_{jj} > 0.4$ and from the reconstructed photons by 
$\Delta R_{\gamma j} > 0.4 $.
\end{itemize}
With these choices of final state selection criteria we now proceed to select the events for our analysis.
%%%%%%%%%%%%%%%%%%%%%%%%%%%%%%%%%%%
\subsection*{Signal Region: $\geq 1$ $\gamma$ $+$ $>$ 2 jets $+$ $\met$}
\label{sec:sigreg}
%%%%%%%%%%%%%%%%%%%%%%%%%%%%%%%%%%%
We look into final states with at least 1 photon, multiple jets and large $\met$. Amongst the existing analyses 
for the same final state carried by the experimental collaborations, the ATLAS analysis imposes a more stringent 
constraint on the new physics parameter space and hence we have implemented the same set of cuts as enlisted 
below for our analysis: 
\begin{itemize}
\item {\bf A1}: The final state events comprise of at least one photon and the leading photon ($\gamma_1$) must have 
$p_T^{\gamma_1}>$ 400 GeV.  
\item {\bf A2}: There should be no charged leptons in the final state ($N_{\ell}$=0) but at least 2 hard jets ($N_j>2$).
\item {\bf A3}: The leading and sub-leading jets must be well separated from $\vec\met$, such that $\Delta\phi(j,\vec\met) >  0.4$.
\item {\bf A4}: The leading photon must also be well separated from $\vec\met$ with $\Delta \phi(\gamma_1,\vec\met) > 0.4$.
\item {\bf A5}: As the light gravitinos would carry away a large missing transverse momenta, we demand  that $\met >  400$ GeV.
\item {\bf A6}: We further demand effective mass, $M_{Eff} > $ 2000 GeV, with $M_{Eff} = H_T + G_T + \met$, 
where $H_T = \Sigma_{i} \text{ } p_T (j_i)$ is the scalar sum of $p_{T}$ of all jets and 
$G_T = \Sigma_j\text{ }  p_T(\gamma_{j})$ is the scalar sum of $p_{T}$ of all photons in the event. 
\end{itemize}
%%%%%%%%%%%%%%%%%%%%%%%%%%%%%%%%%%%%%%
In Table~\ref{tab:sig_nllnlo} we have summarised the effect of the cuts {\bf A0}-{\bf A6} for our signal on the 
respective benchmark points. All the production cross-sections in the table is scaled using
{\tt NLO+NLL} K-factors obtained from {\tt NLL}$\_${\tt Fast} \cite{Beenakker:1996ch,PhysRevLett.102.111802,PhysRevD.80.095004,1126-6708-2009-12-041,Beenakker:2015rna}.
%%%%%%%%%%%%%%%%%%%%%%%%%%%%%%%%%%%%%%%%
\begin{table}[H]
\begin{center}
\small
\begin{tabular}{|c|c|c|c|c|c|c|c|c|} 
\hline 
\multicolumn{2}{|c|}{Signal} & 
\multicolumn{6}{|c|}{Effective cross-section (in fb) after the cuts } \\
\hline 
Benchmark & Production               
&{\bf A0 $ + $ A1} & {\bf A2} & {\bf A3} & {\bf A4} &  {\bf A5} & {\bf A6}   \\
 Points   &  cross-section(fb)  & &    
        &  &    &    &     \\
\hline 
C1 &0.26&0.22&0.18&0.14&0.14&0.13&0.12\\
C2 &0.80&0.68&0.37&0.30&0.30&0.28&0.26\\
C3 &0.23&0.18&0.10&0.08&0.08&0.08&0.08\\
C4 &0.21&0.15&0.12&0.08&0.08&0.08&0.07\\
C5 &0.49&0.34&0.15&0.13&0.13&0.12&0.11\\
C6 &0.77&0.61&0.13&0.11&0.11&0.10&0.09\\
U1 &0.20&0.10&0.09&0.08&0.07&0.05&0.05\\
U2 &0.20&0.13&0.12&0.10&0.09&0.08&0.08\\
\hline
\end{tabular}
\caption{Signal Cross-sections (NLO+NLL) for all the benchmark points listed in Table~\ref{tab:bp_comp} 
corresponding to ($\geq 1$ $\gamma$ $+$ $>$ 2 jets $+$ $\met$) final state. For all the points, $m_{\widetilde{G}}$ = 1 keV.} 
\label{tab:sig_nllnlo}
\end{center}
\end{table}
%%%%%%%%%%%%%%%%%%%%%%%%%%%%%%%%%%

As evident from Table~\ref{tab:sig_nllnlo}, cut efficiencies vary depending on the $compression$ in the spectra. For example, 
the jet requirement affects the signal cross-section of {\bf C6} the most, since it is the most compressed spectra among all. 
Naturally, one would expect jet multiplicity to be smaller in this case compared to the others. As a result, the requirement 
$N_j>2$ reduces the corresponding signal cross-section by a significant amount, whereas, for the uncompressed spectra, {\bf U1} 
and {\bf U2}, this cut has no bearing. The hard photon(s) in the signal events and the presence of direct source 
of $\met$ ensure that the $\met$ and $M_{Eff}$ cuts are easily satisfied by the selected events. 

For the corresponding background events, we use the observed number of background
events at ATLAS, which is 1, for the same final state studied at an integrated luminosity of 13.3 fb$^{-1}$ at 13 TeV 
\cite{ATLAS:2016fks}. The statistical signal significance is computed using 
\begin{equation*}
 \mathcal{S} = \sqrt{2 [(s+b)\ln(1+\frac{s}{b})-s]}
\end{equation*}
where $s$ and $b$ represent the remaining number of signal and background events after implementing all the cuts. 
In Table~\ref{tab:sign}, we have shown the required integrated luminosity to obtain a 3$\sigma$ and 5$\sigma$ 
statistical significance for our signal corresponding to all the benchmark points. 
%%%%%%%%%%%%%%%%%%%%%%%%%%%%%%%%%%%%
\begin{table}[h]
\begin{center}
\begin{tabular}{|c|c|c|}
\hline
Signal & \multicolumn{2}{|c|}{Luminosity $\mathcal{L}$ (in fb$^{-1}$) for} \\
\cline{2-3}
&  $\mathcal{S}$ = 3$\sigma$ & $\mathcal{S}$ = 5$\sigma$ \\
\hline
C1 &  68 & 189\\
C2\tablefootnote{ On the face of it, this benchmark may be ruled out by the current searches
at LHC. However, this is to be taken with some caution, since the search criteria suggested by us 
are slightly different from the ones used in the current experimental searches.} &  19 & 52 \\
C3 &  139 & 385 \\
C4 & 176 & 489\\
C5 & 79 & 219\\
C6 & 112 &312\\
U1 & 326& 904\\
U2 &139 & 385\\
\hline
\end{tabular}
\caption{Required luminosity (\textit{$\mathcal{L}$}) to obtain 3$\sigma$ and 5$\sigma$ statistical Significance \textit{($\mathcal{S}$)} of the 
signal at the 13 TeV run of the LHC corresponding to the benchmark points.}
\label{tab:sign}
\end{center}
\end{table}
%%%%%%%%%%%%%%%%%%%%%%%%%%%%%%%%%%%%

The required luminosity for 3$\sigma$ and 5$\sigma$ statistical significance varies depending 
on the relative compression and heaviness of the spectra. As evident, {\bf C2} has the best discovery 
prospects and is likely to be probed very soon. {\bf C6} on the other hand, despite of having a similar 
squark-gluon spectra and a very similar production cross-section to that of {\bf C2}, requires a much 
larger luminosity ($\sim 112~{\rm fb}^{-1}$) to be probed. This is because the high amount of  compression in the spectra 
reduces the cut efficiency significantly due to the jet multiplicity requirement. The required integrated 
luminosity for {\bf C1} and {\bf C5} is very similar although {\bf C5} has a relatively lighter 
coloured sector and thus a larger production cross-section compared to {\bf C1}, as can be seen 
from Table~\ref{tab:sig_nllnlo}. However, the photon and jet selection criteria reduces the {\bf C5} 
cross-section making it comparable to that of {\bf C1}. 
The situation is different for {\bf U1} which despite of having the lightest gluino, requires the largest luminosity 
($\sim 326~{\rm fb}^{-1}$) among all the benchmark points in order to be probed. The reason is two-fold. Firstly, the 
production cross-section in this case (and also for {\bf U2}) is comprised of just the gluino-pair since the squarks are far 
too heavy to contribute. Secondly, the $\lspone$ being $\sim$ 700 GeV, the photons arising from $\lspone$ decay are 
relatively on the softer side and hence the photon selection criteria further reduces the signal cross-section. 
A similar squark-gluon spectra in presence of a heavier $\lspone$ ({\bf U2}) therefore is likely to be probed with a much 
smaller luminosity ($\sim 139~{\rm fb}^{-1}$) than {\bf U1}. Thus it is evident from 
Table~\ref{tab:sig_nllnlo} and \ref{tab:sign}, that given the present experimental constraints, 
a compressed spectra, unless it is too highly compressed such that the cut efficiency is reduced 
significantly, can improve the squark-gluino mass limits by a significant amount. For example, 
{\bf C2} can be probed with slightly little more luminosity than 13.3 ${\rm fb}^{-1}$ but with 
a coloured spectra that lies in the vicinity of 2.5 TeV. This clearly suggests that a compressed spectra becomes 
much more quickly disfavoured over an uncompressed spectra with a gravitino LSP contrary to the case where a 
compressed SUSY spectrum appears as a saviour of low mass SUSY with a neutralino LSP. This is because of the hard 
photons that themselves act as a clear criterion to distinguish the signal over the SM background.   
%%%%%%%%%%%%%%%%%%%%%%%%%%%%%%%%%%%%
\subsection{Distinction of Compressed and Uncompressed spectra}
\label{sec:distcomp}
%%%%%%%%%%%%%%%%%%%%%%%%%%%%%%%%%%%%
Given the inclusive hard photon + $\met$ signals, supposedly due to a light gravitino, can one ascertain whether 
the MSSM part of the spectrum is compressed or uncompressed? With this question in view, it is worthwhile 
to compare signals of both types with various degree of compression in presence of a light ($\sim$ keV) gravitino 
as the LSP. We show that the kind of compressed spectra 
we have used enhances the existing exclusion limit on the coloured sparticles. We consider different squark-gluino 
mass hierarchy represented by our choice of some sample benchmark points
presented in Table~\ref{tab:bp_comp}. The $\grav$ being almost massless in comparison to the $\lspone$ in consideration, 
the photons generated from the $\lspone$ decay into $\grav$ are always expected to be very hard for both the compressed 
and uncompressed scenarios. This feature can be used to enhance the significance of the signal irrespective of the associated 
jets in the event. We provide a framework where one can use the properties of these jets in a novel way to distinguish 
between the two different scenarios in consideration even if they produce a similar event rate at the LHC. 
For illustration, let us consider the benchmark points, {\bf C5}, {\bf C4} and {\bf U2} all of which 
result in nearly identical event rates for our signal and thus
it is difficult to identify whether it is a signature of a compressed or an uncompressed spectra.
It would be nice to have some kinematic variables which could be used to distinguish among 
the different kind of spectra. Subsequently, we have proposed few such variables which show 
distinctive features in their distributions depending on the relative hardness and multiplicity 
of the final state photon(s) and jets.   

\begin{figure}[H]
\includegraphics[scale=0.28]{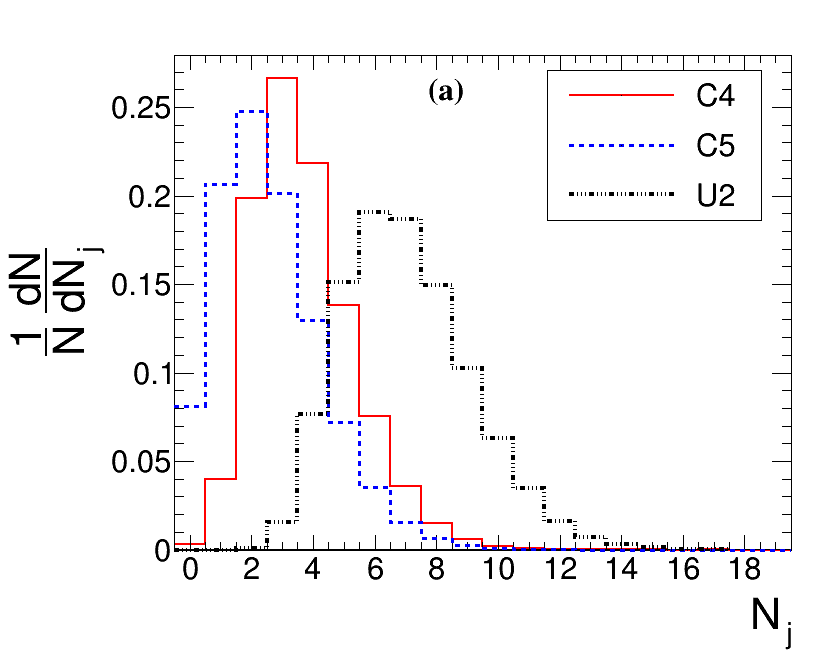}
\includegraphics[scale=0.285]{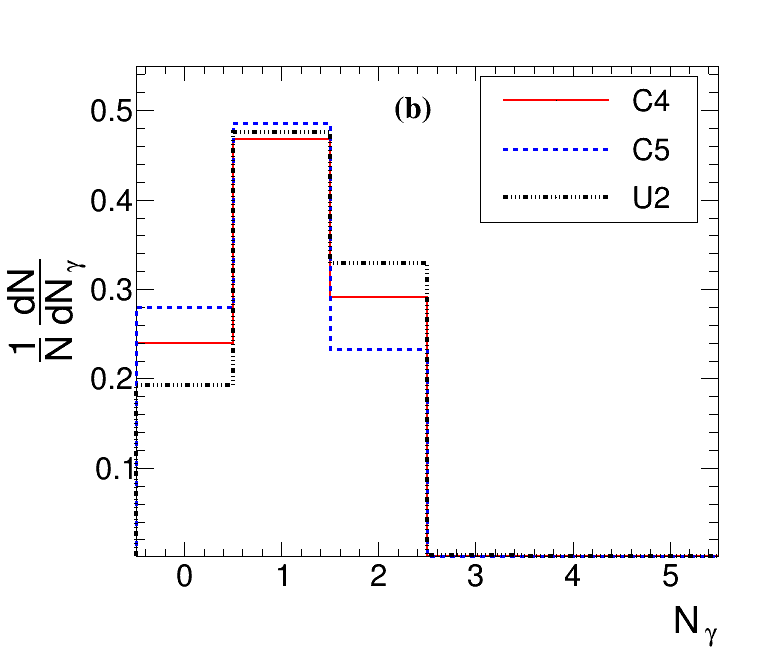}\\
\caption{Normalized distributions for jet and photon multiplicity for the benchmark points 
{\bf C4}, {\bf C5} and {\bf U2} representing moderately compressed, highly compressed and uncompressed scenarios respectively. 
Figure (a) has been prepared after implementing the selection cuts {\bf A0+A1} and figure (b) after {\bf A0}.}
\label{fig:dist_var1}
\end{figure}

An uncompressed spectrum, such as {\bf U2} is characterized by a large mass gap between the strong sector sparticles 
and the NLSP ($\lspone$). This ensures a large number of high $p_{T}$ jets from the cascades as compared to {\bf C5} 
and {\bf C4}. The difference in jet multiplicity in the two cases is clearly 
visible in Fig.~\ref{fig:dist_var1} where we have presented both the jet and photon multiplicity 
distributions for some sample compressed and uncompressed spectra. The hard photons in the event are originated from the 
$\lspone$ decay and since for all our benchmark points the $\lspone$ is sufficiently heavy, the photon multiplicity peaks at a 
similar region for both the compressed and uncompressed spectra. However, the jets in the case of {\bf U2} are generated from 
the three body decay of the gluino into a pair of quarks and $\lspone$. As evident from Fig.~\ref{fig:gl_br}, for the choices of 
the sparticle masses of {\bf U2}, the other decay mode is highly suppressed. Hence one would naturally expect to obtain 
a large number of jets in the final state as shown in Fig.~\ref{fig:dist_var1}. {\bf C5} having a high degree of compression 
($\Delta M_i = 48$ GeV) in the parameter space results in least number of jets in the final state.  
{\bf C4}, on the other hand, has a more relaxed compression ($\Delta M_i = 198$ GeV) that gives rise to slightly harder 
cascade jets passing through the jet selection criteria resulting in a harder distribution than {\bf C5}. 
%%%%%%%%%%%%%%%%%%%%%%%%%
%%%%%%%%%%%%%%%%%%%%%%%%%

\begin{figure}[ht]
\includegraphics[scale=0.265]{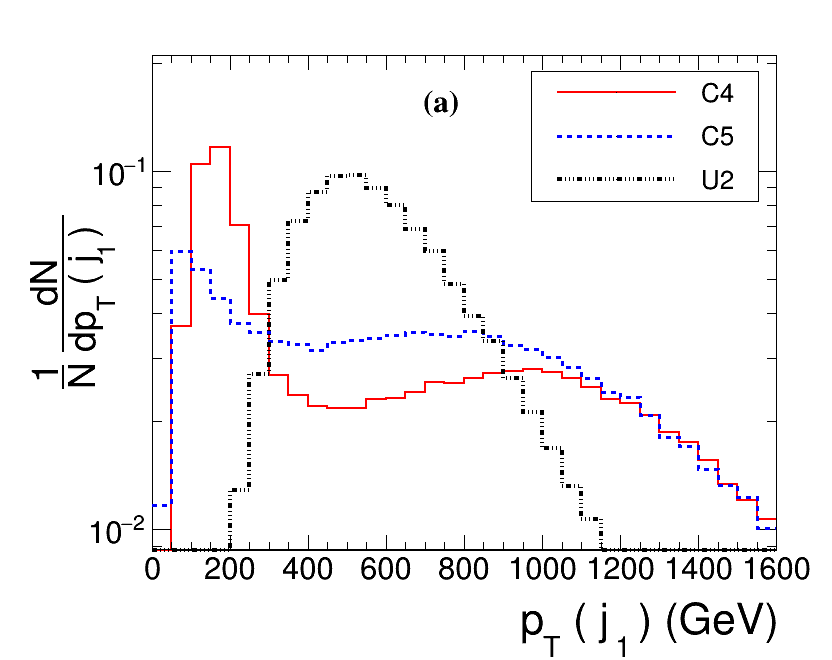}
\includegraphics[scale=0.26] {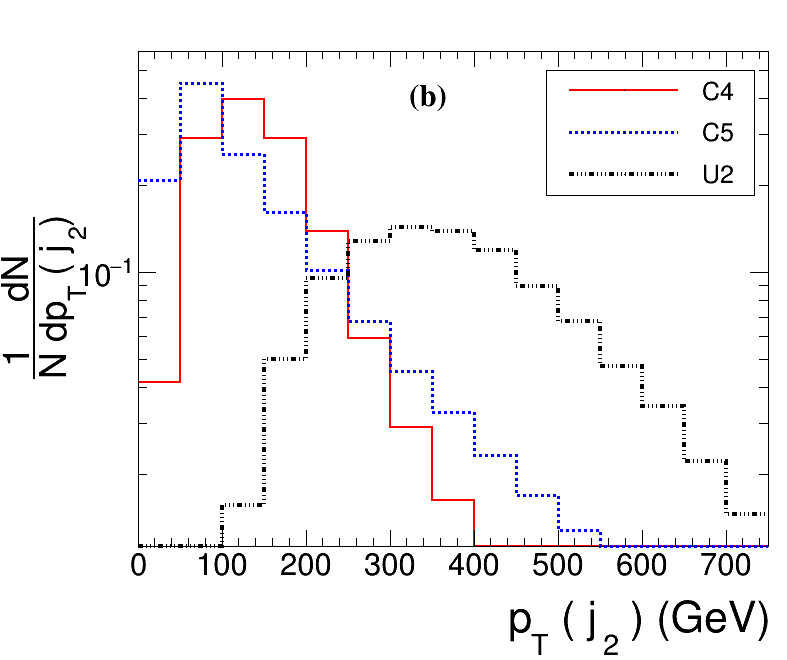} \\
\includegraphics[scale=0.26]{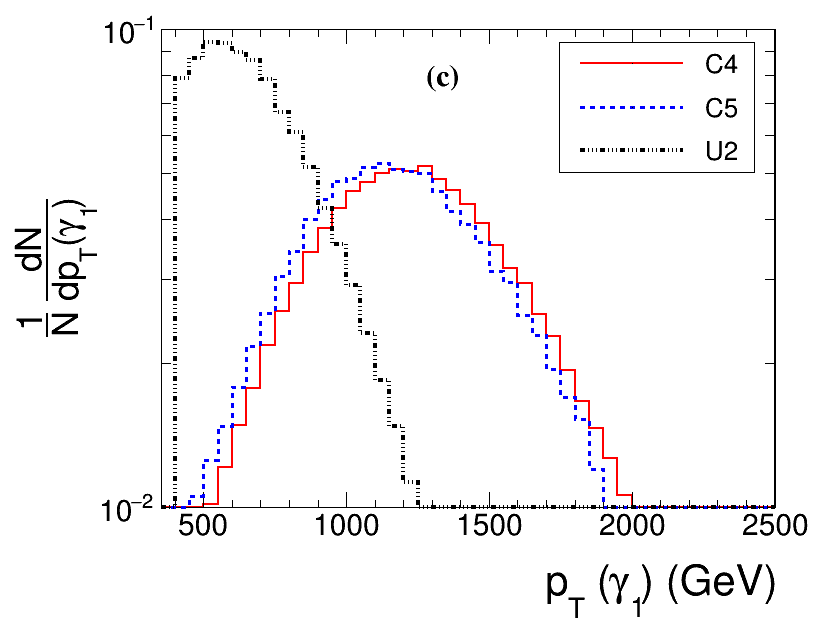}
\includegraphics[scale=0.28]{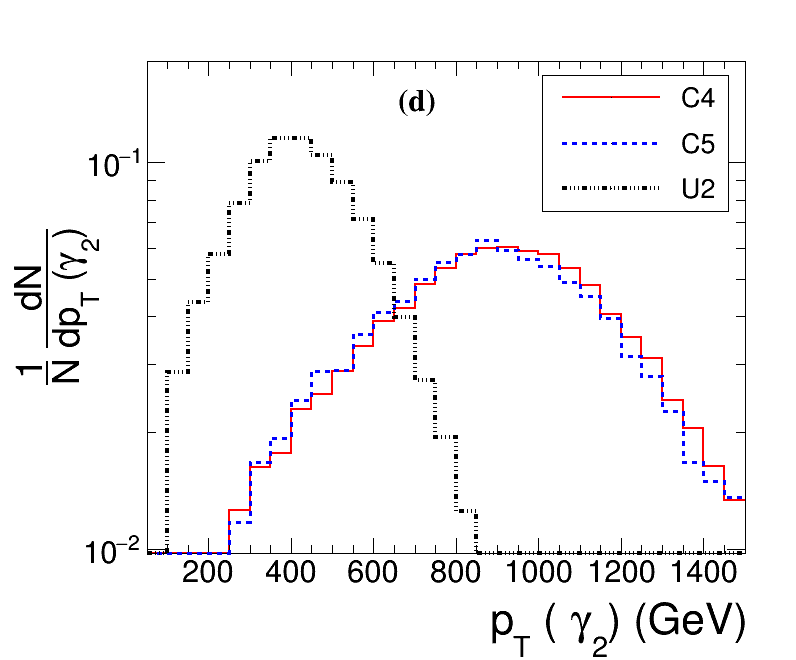}\\
\caption{The leading and subleading jet and photon $p_{ T}$ distributions for some of the benchmark points representing various compressed ({\bf C4}), more 
compressed ({\bf C5}) and uncompressed ({\bf U2}) spectra after implementing the selection and analysis cuts {\bf A0}-{\bf A6}.}
\label{fig:dist_var2}
\end{figure}

The relative difference in the compression factor ($\Delta M_i$) among the three benchmark points are also visible in the 
jet $p_{T}$ distributions shown in Fig.~\ref{fig:dist_var2}. 
As expected, the leading (Fig.~\ref{fig:dist_var2}(a)) and subleading (Fig.~\ref{fig:dist_var2}(b)) jet $p_{T}$ 
distributions predominantly show a harder peak
for {\bf U2} as compared to {\bf C4}, {\bf C5}. 
However, hard jets may also arise from the $\lspone$ decaying to a $Z$ boson and gravitino 
(BR $\sim$ 25$\%$) as the $Z$ decays dominantly into two jets. The $Z$ boson is expected to be highly boosted 
and thus one can easily obtain additional hard jets from its decay. These jets populate a small fraction of the total number of 
events and thus for a compressed spectra one of these jets can turn out to be the hardest jet in the event. This feature can 
be observed by the subdominant peak at $\sim$ 1000 GeV for the leading jet $p_{T}$ distribution in Fig.~\ref{fig:dist_var2}. 

\begin{figure}[ht]
\includegraphics[scale=0.285]{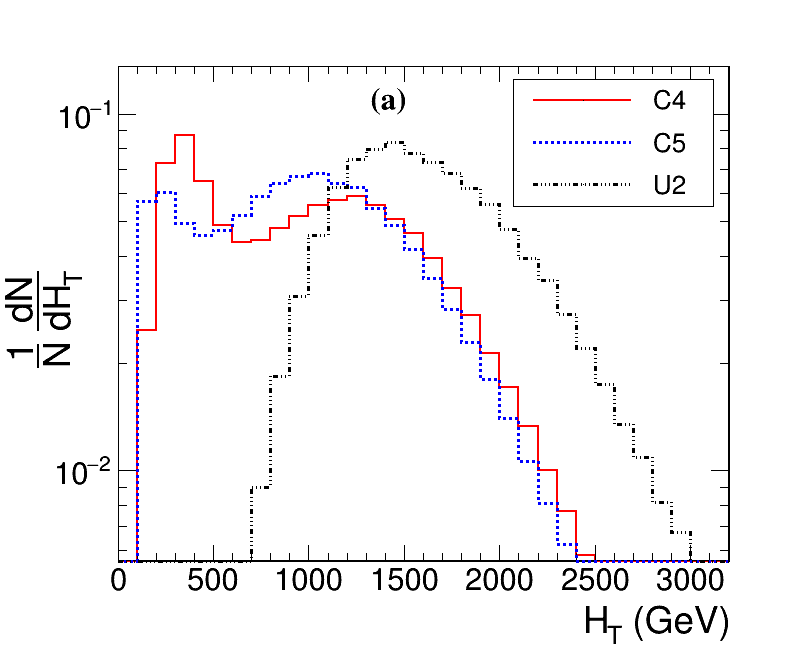}
\includegraphics[scale=0.295]{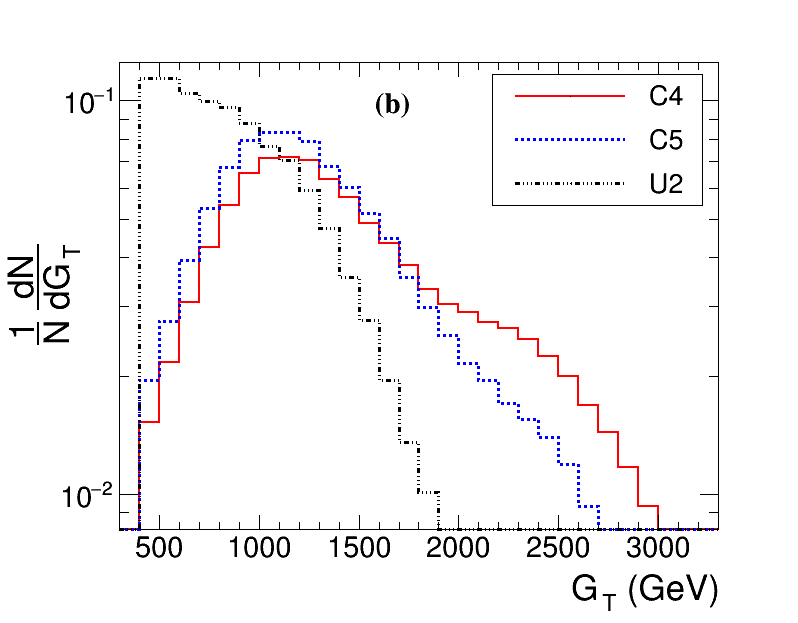}\\
\includegraphics[scale=0.29]{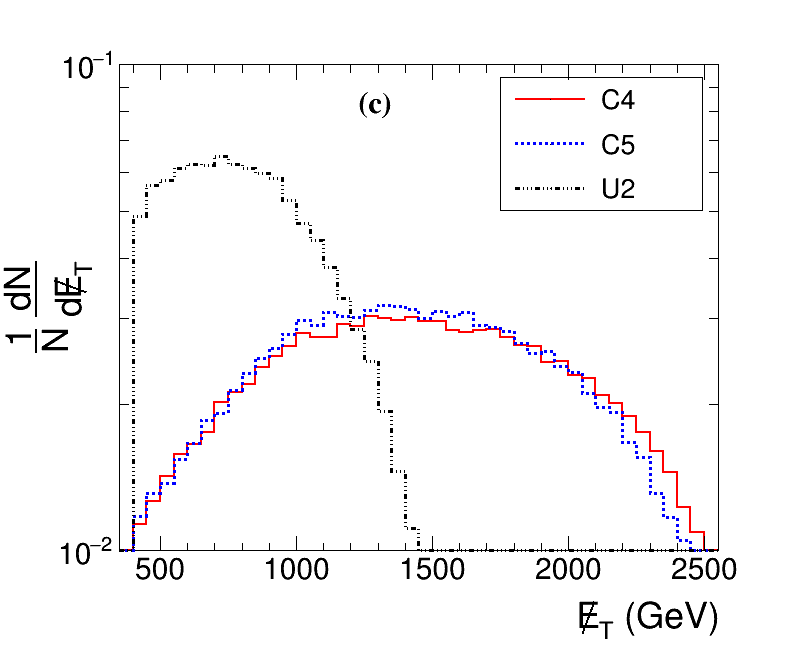}
\includegraphics[scale=0.29]{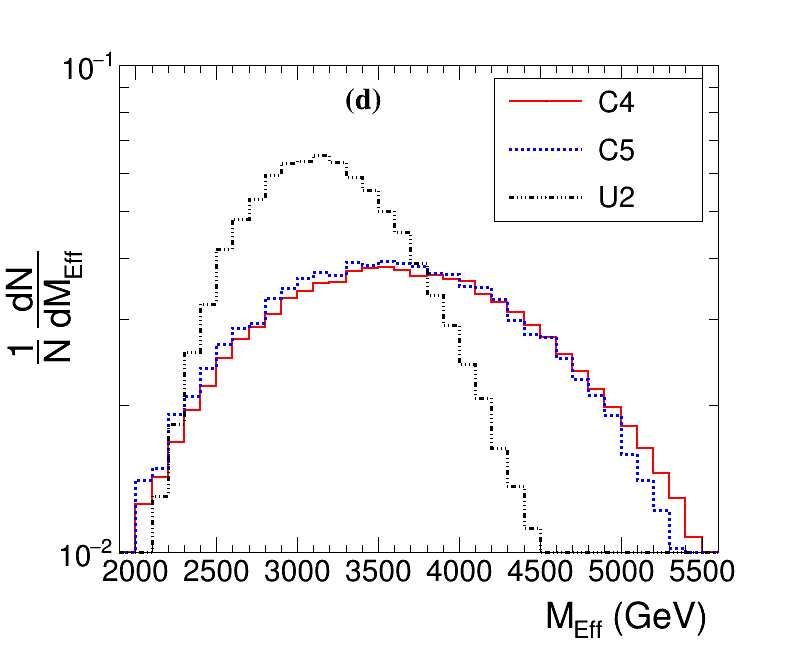}\\
\caption{Normalized distributions of total hadronic energy deposit $H_T$,total photon energy deposit $G_T$, missing transverse energy 
$\met$ and Effective Mass $M_{Eff}$, for benchmark points representing various compressed ({\bf C4}), more compressed ({\bf C5}) and uncompressed ({\bf U2}) spectra after 
implementing the selection and analysis cuts {\bf A0}-{\bf A6}.}
\label{fig:dist_var3}
\end{figure}

Fig.~\ref{fig:dist_var2}(c) and (d) show the leading and subleading photon $p_{T}$ distributions respectively 
for {\bf C4}, {\bf C5} and {\bf U2}. 
The $\lspone$ mass in {\bf C4}, {\bf C5} being $\sim$ 2.5 TeV, the photons produced from their decay are much harder than the 
leading jets in the spectra as opposed to the uncompressed spectra ({\bf U2})
% for which the NLSP mass is kept at a lower value ($\sim 1200$ GeV) 
and hence, the peak in the photon $p_{T}$ distribution is significantly shifted to lower values. 
Thus while the total hadronic energy, $H_{T}$ (Fig.~\ref{fig:dist_var3}(a)) peaks at a higher value for the uncompressed 
case owing to a large number of hard jets, $G_{T}$ (Fig.~\ref{fig:dist_var3}(b)) which is the scalar sum of all photon 
$p_{T}$, peaks at a lower value for the uncompressed case than the compressed cases. 
%%%%%%%%%%%%%%%%%%%%%%%%%%%%
%%%%%%%%%%%%%%%%%%%%%%%%%%%%%%

Among other kinematic variables, one can also look into the $\met$ and M$_{Eff}$ distributions 
to distinguish the compressed and uncompressed scenarios as shown in Fig.~\ref{fig:dist_var3}(c) and (d) respectively.
Since the photons are almost always harder for the compressed spectra compared to the uncompressed cases, we have 
observed that the $\met$, required to balance the total visible transverse energy, is much harder for the former. 
Effective mass, M$_{Eff}$ defined as the sum of $H_{T}$, $G_{T}$ and $\met$, also shows some small difference in the 
peak value for both cases. In {\bf U2}, $G_{T}$ and $\met$ are softer than that for {\bf C4}, {\bf C5} but $H_T$ is much harder 
resulting in the $M_{Eff}$ peaking at similar values for the both cases.
However, since the photons are considerably harder than the jets in all cases, 
the effect being more pronounced for the compressed over the uncompressed case,
the M$_{Eff}$ distribution falls faster for {\bf U2} than {\bf C4} and {\bf C5} as can be seen 
from Fig.~\ref{fig:dist_var3}(b) and \ref{fig:dist_var3}(d) respectively.

\begin{figure}[H]
\includegraphics[scale=0.27]{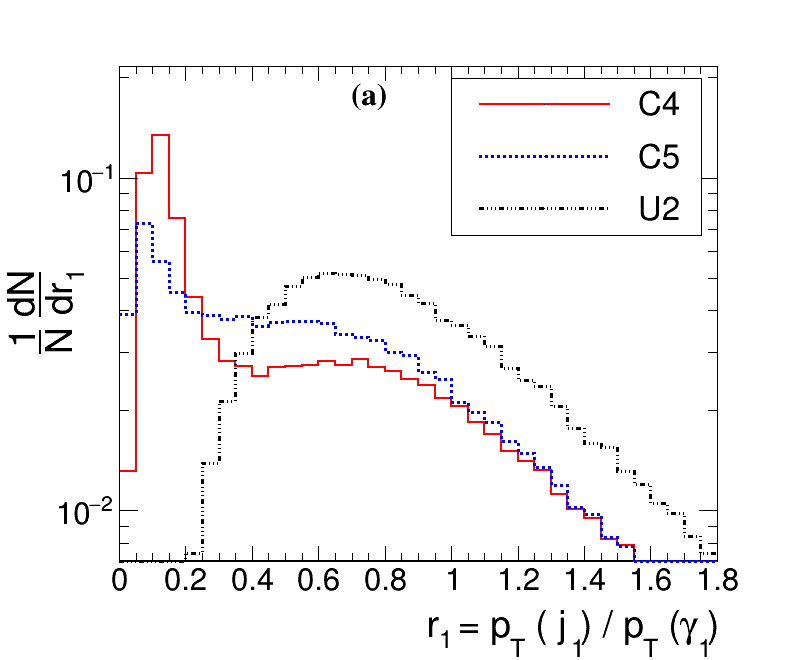}
\includegraphics[scale=0.277]{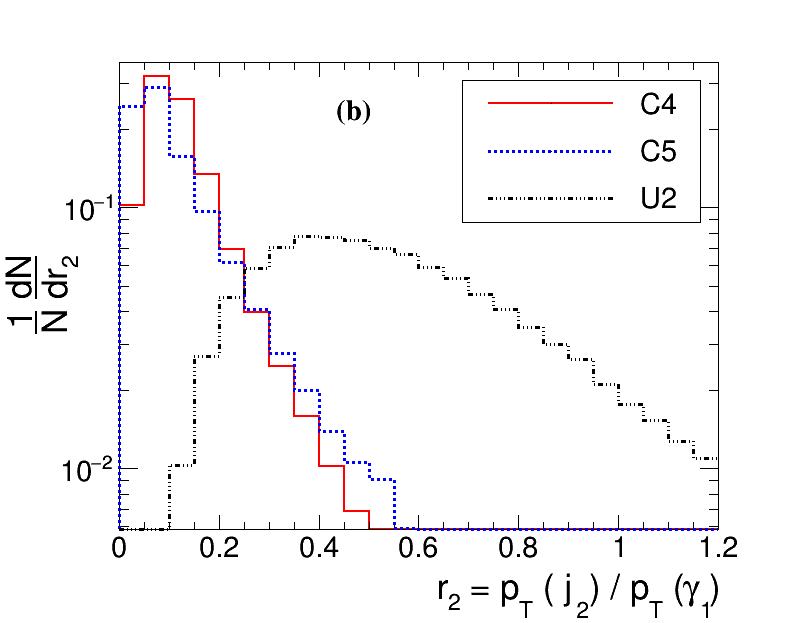}\\
\includegraphics[scale=0.27]{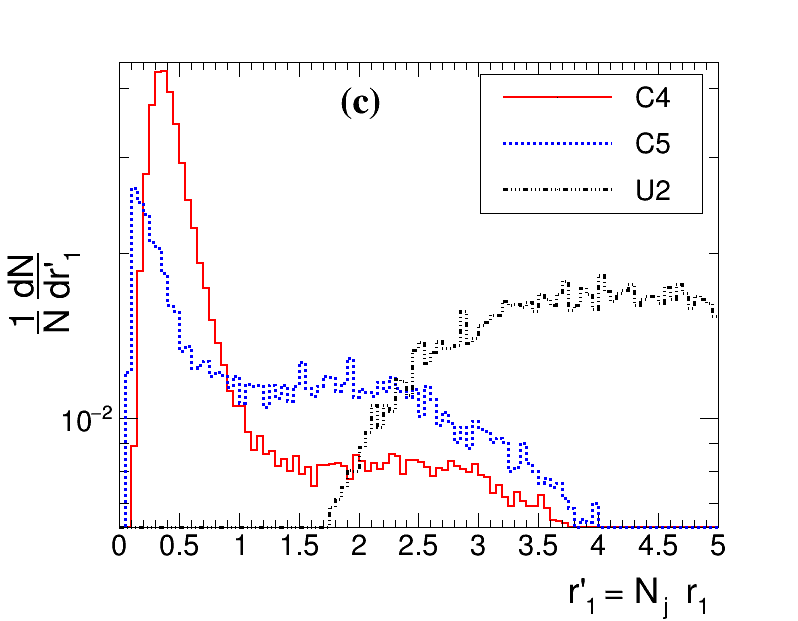}
\includegraphics[scale=0.275]{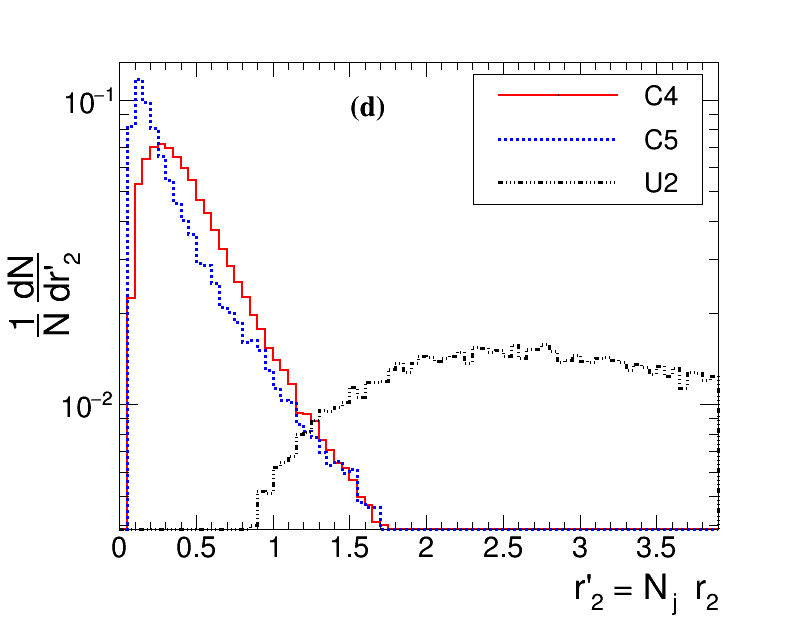}\\

\caption{Normalized distributions of different kinematic variables $r_{1}$, $r_{2}$, $r_{1}^{\prime}$ and $r_{2}^{\prime}$ to distinguish compressed and uncompressed scenarios for some of the benchmark points 
representing various compressed ({\bf C4}), more compressed ({\bf C5}) and uncompressed ({\bf U2}) spectra
after implementing the selection and analysis cuts {\bf A0}-{\bf A6}.}
\label{fig:dist_var4}
\end{figure}

Taking cue from the kinematic distributions in Fig.~\ref{fig:dist_var2} and Fig.~\ref{fig:dist_var3}, we now proceed to formulate two observables 
\begin{align*}
r_1 = \frac{p_T(j_1)}{p_T(\gamma_1)}  \,\,\,\,\,\, {\rm and} \,\,\,\,\,\,   
r_2 = \frac{p_T(j_2)}{p_T(\gamma_1)} 
\end{align*}
which capture the essence of the jet and photon transverse momenta behaviour in a way 
as to distinctly distinguish between the compressed and uncompressed scenarios. As seen in Fig.~\ref{fig:dist_var4}, for 
the compressed case, $r_1$ (Fig.~\ref{fig:dist_var4}(a)) peaks at rather small values
($\sim 0.1$ ) than the uncompressed case ($\sim1.0$) since 
the leading jet $p_T$ is almost always softer than the leading photon for compressed spectra whereas for the uncompressed case 
there are hard jets with $p_{T}$ values comparable to the leading photon $p_T$. However for the compressed spectra, 
the collimated hard jet from the highly boosted $Z$ boson produced in the decay of the $\lspone$, lead to a subdominant peak 
at $\sim 0.7$ in $r_{1}$. The observable $r_2$ (Fig.~\ref{fig:dist_var4}(b)) constructed with the sub-leading jet and 
leading photon $p_T$, peaks at 
lower values ($\sim0.1$) for {\bf C4} and {\bf C5} since the sub-leading jet, coming from the cascades 
or ISR in the compressed case is expected to be much softer than the photon. For {\bf U2}, $r_2$ peaks at $\sim0.5$ since the 
sub-leading jet also coming from the cascade is softer than the hardest photon. Thus we find that the above ratios seem to enhance 
the two major distinctive features between a compressed and an uncompressed scenario, namely the high/low $p_T$ for the 
photon/jet for the compressed as compared to the low/high $p_T$ of the photon/jet for the uncompressed case. 

We further note that the jet multiplicity is another variable which shows a 
difference in the distributions for compressed spectra \textbf{C4} and \textbf{C5} when 
compared to that of the uncompressed spectra \textbf{U2} (Fig.~\ref{fig:dist_var1}(a)). Although the choice of our signal region involves $N_{j} > $ 2,
the compressed spectra, \textbf{C4} and \textbf{C5}, still retain a sufficient fraction of events with higher number of jets. 
In contrast, the uncompressed spectra \textbf{U2} has larger number of hard jets for all 
events, and thereby remains mostly unaffected by this selection criterion. We therefore
define a modified ratio (scaled by the jet multiplicities) as 
\begin{align*}
  r_{1}^{\prime} = N_{j} \text{ } r_{1}  \,\,\,\,\,\, {\rm and} \,\,\,\,\,\, r_{2}^{\prime}= N_{j} \text{ } r_{2}.
\end{align*}
 Notably the new variables $r_{1}^{\prime}$ and $r_{2}^{\prime}$ are able to significantly 
 enhance the differences between a compressed and uncompressed spectra. 
Since the scale factor, $N_{j}$, is always greater for the uncompressed spectra 
 \textbf{U2} than for the compressed spectra \textbf{C4} and \textbf{C5}, 
we find the peak values of $r_{1}^{\prime}$ ($\sim$ 4.0) and $r_{2}^{\prime}$ ($\sim 2.5$) of the uncompressed spectra are shifted further away from that of 
compressed ones ($r_{1}^{\prime}$ $\sim$ 0.2-0.5 and $r_{2}^{\prime}$ $\sim$ 0.1-0.3). Quite importantly the visible overlap seen in $r_1$ for the sub-dominant peak is now completely disentangled in the new variable $r_{1}^{\prime}$ as seen in Fig.~\ref{fig:dist_var4}(c). 
This is significant in the sense that when the event samples would retain a much 
harder criterion for the leading jet then the events for \textbf{U2}, \textbf{C4} and 
\textbf{C5} would all feature the overlap observed 
for the sub-dominant peak while the difference for low $r_1$ might be washed away 
for this particular choice of event selection.

%%%%%%%%%%%%%%%%%%%%%%%%%%%%%%%
Besides enhancing the differences between the compressed and uncompressed spectra, the differential distributions in $r_{i}$ and $r_{i}^{\prime}$ can also be used to 
highlight the differences amongst the different compressed spectra themselves, depending 
on the level of compression in mass.
For example, \textbf{C4}, has a larger mass separation $\Delta M_{i}$ than \textbf{C5}, and 
shows a peak in the jet multiplicity at $N_{j}$ = 3 while for \textbf{C5}, the peak value of the
differential cross section is at $N_{j}$ = 2. Thus a larger 
fraction of events survive after analysis for \textbf{C4} than \textbf{C5}. 
Again, since \textbf{C5} is relatively more compressed than \textbf{C4}, the jets from \textbf{C4} are considerably harder than the latter.
However the NLSP mass for \textbf{C4} is larger than \textbf{C5}, since to probe lower values of compression, we require a heavier NLSP to meet current LHC bounds.
This results in the photons being harder for \textbf{C4} than for \textbf{C5}. 
The combined effect of the two seem to be more prominent for both $r_{1}$ and $r_{1}^{\prime}$, where the leading jet is either the ISR jet or 
cascade jet in case of \textbf{C4}.
For $r_{2}$ this effect seems neutralised, owing to the sub-leading jets for both cases, being much softer than the leading photon $p_T$.
However the scale factor $N_{j}$ shifts the peak value of $r_{2}^{\prime}$, thus efficiently distinguishing amongst the two compressed spectra of varying degree of compression.
%%%%%%%%%%%%%%%%%%%%%%%%%%%%%%%%
\subsection{eV Gravitino}
\label{sec:evgr}
%%%%%%%%%%%%%%%%%%%%%%%%%%%%%%%%
As pointed out earlier that the kinematic characteristics of events when the NLSP decays into a gravitino 
are independent of whether the $\grav$ is in the keV or eV range. Therefore, for an NLSP decaying into a $\grav$ and a SM 
particle, the $\grav$ is practically massless. However, as discussed
in Section~\ref{sec:spbr}, a lighter gravitino has a stronger coupling strength to the sparticles. Thus 
the decay of the sparticles into a SM particle and gravitino dominates over 
its decay to the NLSP. For a gravitino of mass 1 eV, we find that the gluino/squark  almost always  directly  decays to 
the gravitino rather than to the NLSP. The branching fractions also depend on the
mass gap between the coloured sparticles and the NLSP. 
These features are highlighted in Figs. \ref{fig:gl_br} and \ref{fig:sq_br} where both compressed and uncompressed
mass gaps are shown. 

Therefore, an eV $\grav$ does affect the overall event rates of the signal in the photon channel when compared to the 
keV $\grav$ case. An immediate consequence which has gone unnoticed for such light eV $\grav$ case would be a new competing 
signal which can become more relevant than the more popular photonic channel. This can be easily understood by taking a look at the 
resulting BR($\widetilde{g}\rightarrow g \grav$) for some of our benchmark points in presence of an eV gravitino. As indicated by 
Fig. \ref{fig:gl_br}, this branching ratio is supposed to go up if the spectrum is more compressed.
For the same benchmark points as in Table~\ref{tab:bp_comp}, now in the presence of an eV $\grav$, we have observed that
BR($\widetilde{g}\rightarrow g \grav$) $\sim 13\%, 41\%$ and $99\%$ for  \textbf{U1}
($\Delta m_{\gl\lspone}$ = 1403 GeV), \textbf{U2} ($\Delta m_{\gl\lspone}$ = 911 GeV)
and \textbf{C1} ($\Delta m_{\gl\lspone}$ = 78 GeV) respectively.
As a consequence, \textbf{C1} with an eV gravitino, is unlikely to yield a good event rate in the photonic channel since 
the gluino avoids decaying into the NLSP altogether. However, a small fraction of the squarks may still decay into the NLSP, $\sim 4\%$ and 
$\sim 24\%$ precisely for left and right squarks respectively. Hence, one would still expect a photon signal for such a scenario, but a much 
weaker one as presented in Table~\ref{tab:c1grav}. 

%%%%%%%%%%%%%%%%%%%%%%%%%%%%%%%
\begin{table}[H]
\begin{center}
\small
\begin{tabular}{|c|c|c|c|c|c|c|c|}
\cline{3-8}
\hline
Signal & Production  &\multicolumn{6}{|c|}{Cross-section (in fb) after cuts$:$}\\ 
\cline{3-8}
 &  cross-section (in fb) & \bf{A0+A1}& \bf{A2}& \bf{A3}&\bf{A4} &\bf{A5} &\bf{A6} \\
\hline
C1 &0.26&0.038&0.035&0.031&0.03&0.028&0.028 \\
\hline
\end{tabular}
\caption{Signal Cross-sections (NLO+NLL) for benchmark {\bf C1} for $\geq$ 1 photon $+$ $>$ 2 
jets $+$ $\met$ final states ($m_{\widetilde{G}}$ = 1 eV).}
\label{tab:c1grav}
\end{center}
\end{table}
%%%%%%%%%%%%%%%%%%%%%%%%%%%%%%%
As expected,  the photon signal weakens considerably  
when compared to one with a keV gravitino and requires an integrated luminosity $\sim$ 1000 fb$^{-1}$ for observation at the LHC.
However, much stronger signal would be obtained in the ``n-jet+$\met$'' ($n\ge$2) channel as the final state would have at least two very 
hard ($p_T$'s exceeding more than a TeV) jets and an equally hard $\met$ signal for the eV-gravitino case. 
The conventional multi-jet search \cite{ATLAS:2016kts} rely upon the usual $\met$, $M_{Eff}$, 
$\frac{\met}{\sqrt{H_T}}$ and $\Delta\phi(j,\vec{\met})$ cuts and in some cases, razor variables \cite{CMS:2016nnn} to reduce the 
SM backgrounds. We have checked that with these cuts, a 3$\sigma$ significance can be achieved for {\bf C1} in the ``n-jet+$\met$'' 
($n\ge$2) final state at an integrated luminosity of $\sim$ 1000 ${\rm fb}^{-1}$. However, in the presence of an eV gravitino, 
one can demand harder $p_T$ requirements of the jets and harder $\met$, $M_{Eff}$ along with the other conventional cuts to increase 
signal significance further. We have checked that one can easily bring down the required luminosity to $\sim$ 728 ${\rm fb}^{-1}$ 
for a 3$\sigma$ significance, which is a big improvement over the results obtained for the photon-associated 
final state. Thus the multi-jet channel is the more favorable one in order to explore an eV gravitino in presence of a $\sim$ TeV 
compressed colour sector. However, as mentioned earlier, such a light gravitino may 
not be a viable dark matter candidate and would necessarily require the presence of other candidates to satisfy the constraints.

%%%%%%%%%%%%%%%%%%%%%%%%%%%%%%%%%%%%%%%%
\section{Summary and Conclusion}
\label{sec:sumcon}
%%%%%%%%%%%%%%%%%%%%%%%%%%%%%%%%%%%%%%%%
In this work, we have explored the compressed SUSY scenario in the presence of a light gravitino LSP within the framework 
of phenomenological MSSM. The question asked is: since the light gravitino produced in the (neutralino) NLSP decays
generates as much $\met$ for compressed spectra as for uncompressed ones, 
are the former discernible?

The existing collider studies for such scenarios mostly account for the uncompressed parameter regions, and in some cases 
the NNLSP-NLSP compressed regions. However, compression in the entire coloured sector of the sparticle spectrum can result 
in significantly different exclusion limits on the masses of squark, gluino and the lightest neutralino. The presence of a light gravitino in 
the spectrum affects the branching ratios of the coloured sparticles into $\lspone$. We have studied the interplay of these relevant 
branching ratios for varying $\grav$ mass and different amount of compression in the rest of the sparticle spectrum for a bino-like 
$\lspone$. Dictated by the DM constraints, we have mostly concentrated on the keV $\grav$ scenario and have performed a detailed 
collider simulation and cut-based 
analysis for $\geq$ 1 photon + $>$ 2 jets $+$ $\met$ final states arising from the squark-gluino pair production channels in the 
context of the LHC. In our case, the squarks and the gluinos dominantly decay into the $\lspone$ which further decay into a $\grav$ 
along with a $\gamma$ or a $Z$ resulting in the above mentioned final state. Hard $p_T$ photon requirement can be used 
along with other kinematic cuts to 
suppress the SM background very effectively. We have followed the existing ATLAS analysis for the same 
final state with the help of some benchmark points. We have shown that with the existing experimental data, 
the exclusion limits on the coloured sparticle masses can increase by $\sim$ 500 GeV for a highly compressed sparticle spectra. It is 
understood that similar signal event rates can be obtained from both uncompressed and compressed spectra depending on the choices 
of masses of squark, gluino and the lightest neutralino. However, the difference in the compression will be reflected in the kinematic 
distributions of the final state jets and photons. We have exploited this fact to construct some variables which can be used to good effect 
to differentiate between the two scenarios. We have also studied the collider prospects of SUSY spectra in the presence of sub-keV gravitinos. 
It turns out that in such cases, the $\grav$-associated decay modes of the heavy ($\sim$ 2.5 TeV) coloured sparticles start to become 
relevant in the presence of high compression between the NNLSP and NLSP. Then the most suitable final state to look for such spectra would be 
multi-jets + $\met$. However, the existing DM constraints strongly disfavour presence of such light gravitino in the spectrum. 
%%%%%%%%%%%%%%%%%%%%%%%%%%%%%%%%%%%%%%%%
\section{Acknowledgement}
\label{sec:ack}
The work of JD, SM, BM and SKR is partially supported by funding available from
the Department of Atomic Energy, Government of India, for the Regional Centre for
Accelerator-based Particle Physics (RECAPP), Harish-Chandra Research Institute, HBNI.
PK thanks RECAPP for the hospitality during this work.
Computational work for this work was carried out at the cluster computing facility 
in the Harish-Chandra Research Institute (http://www.hri.res.in/cluster).
%%%%%%%%%%%%%%%%%%%%%%%%%%%%%%%%%%%%%%%%
%************************************************************************** NEW SECTION *******************************************************************
\providecommand{\href}[2]{#2}
\addcontentsline{toc}{section}{References}

\bibliographystyle{JHEP}
\bibliography{compsusy_grav}

\end{document}